\documentclass[prd,aps,preprintnumbers,showpacs,
showkeys,superscriptaddress,showpacs,nofootinbib,byrevtex,fleqn]{revtex4-1} 
\usepackage{amsmath,amsfonts,amssymb,amscd,amsxtra,amsthm}
\usepackage{graphicx,wrapfig}
\usepackage{bm}
\usepackage{color}

\begin{document}
\preprint{INHA-NTG-06/2012}
\title{$\phi$ photoprodution with coupled-channel effects}  

\author{Hui-Young Ryu}
\affiliation{Research Center for Nuclear Physics, 
Osaka University, Ibaraki 567--0047, Japan}
\email{hyryu@rcnp.osaka-u.ac.jp}
\author{Alexander I. Titov}
\affiliation{Bogoliubov Laboratory of Theoretical Physics, JINR, Dubna
141980, Russia}
\affiliation{Institute of Laser Engineering, Yamada-oka, Suita, Osaka
565--0871, Japan}
\email{atitov@theor.jinr.ru}
\author{Atsushi Hosaka}
\affiliation{Research Center for Nuclear Physics, 
Osaka University, Ibaraki 567--0047, Japan}
\email{hosaka@rcnp.osaka-u.ac.jp}
\author{ Hyun-Chul Kim}
\affiliation{Department of Physics, Inha University, Incheon 402--751,
Republic of Korea}
\affiliation{School of Physics, Korea Institute for Advanced Study,
Seoul 130--722, Republic of Korea}
\email{hchkim@inha.ac.kr}
\date{\today}
\begin{abstract}
We study $\phi$ photoprodution with various
hadronic rescattering contributions included, in
addition to the Pomeron and 
pseudoscalar meson-exchange diagrams. We find that the hadronic box 
diagrams can explain the recent experimental data in the
vicinity of the threshold. In particular, the bump-like structure at
the photon energy $E_{\gamma} \approx 2.3$ GeV is well explained by
the $K\Lambda(1520)$ rescattering amplitude in the intermediate state, 
which is the dominant contribution among other hadronic contributions.  
We also find that the hadronic box diagrams 
are consistent with the observed spin-density matrix elements near the
threshold region. 
\end{abstract}
\maketitle
\section{Introduction}
The $\phi (1020)$ meson is distinguished from other vector mesons,
since it contains mainly strange quarks. Because of
its dominant strange quark content, its decays to lighter mesons and
coupling to the nucleon are known to be suppressed by the
Okubo-Zweig-Iizuka (OZI) rule. In fact, the strange vector form
factors of the nucleon, which is implicitly related to the $\phi$
meson via the vector-meson dominance, is reported to be rather
small~\cite{Ahmed:2011vp}. This large $s\bar{s}$   
content of the $\phi$ meson makes the meson-exchange picture
unfavorable in describing photoproduction of the $\phi$  
meson. Thus, the Pomeron~\cite{Donnachie_etal,Close_etal} is believed
to be the main contribution to $\phi$ photoproduction, since it
explains the slow rise of the differential cross sections of $\phi$
photoproduction as the energy increases. However, while it is true in
the higher energy region, a recent measurement reported by the LEPS
collaboration~\cite{Mibe:2005er} shows a bump-like structure around the
photon energy $E_\gamma \approx 2.3$ GeV. It seems that the Pomeron
alone cannot account for this bump-like structure and requires that
one should consider other production mechanism of $\phi$
photoproduction near the threshold energy. 
Moreover, a recent measurement of the spin-density matrix
elements near the threshold region~\cite{Chang:2010dg} implies that
hadronic degrees of freedom play essential role in the vicinity of the
threshold. 

So far, the theoretical understanding of the production
mechanism for the $\phi$ photoproduction can be summarized as follows:   
\begin{itemize}
\item General energy-dependence of the cross sections is mainly
  explained by Pomeron exchange that can be taken as 
either a scalar meson or a vector meson with charge conjugation
$C=+1$. While the Pomeron explains the increase of the differential
cross section $d\sigma/dt$ in the forward direction, it cannot
describe the behavior of $d\sigma/dt$ near the threshold.   
\item The exchange of neutral pseudoscalar mesons ($\pi^0$,
  $\eta$) provides a certain contribution to $d\sigma/dt$ near the
  threshold but it is not enough to explain the threshold behavior of
  $d\sigma/dt$~\cite{Titov:1999eu}. Moreover, $\pi^0$ and $\eta$
  exchanges wrongly predict the spin-density observables and, in
  particular, $\rho_{1-1}^1$ matrix element (see Appendix for its
  definition).  
\item Usual vector meson-exchanges such as $\rho$ and $\omega$ are
  forbidden due to their negative charge conjugations 
  ($C=-1$). Otherwise, the charge conjugation symmetry will be broken.   
\item Vector meson-exchanges with exotic quantum number such as
  $I(J^{PC})=1(1^{-+})$ are allowed but those vector mesons are not
  much known experimentally. Moreover, as for the experimental data
  from the deuteron target, exchange of isoscalar mesons is more
  plausible. On the other hand, there is no experimental evidence
  for isoscalar hybrid-exotic 
  mesons~\cite{PDG2012}.   
\item The contribution of scalar mesons such as $\sigma$ and $f_0$ are
  negligibly small for $d\sigma/dt$~\cite{Titov:1999eu}.
\end{itemize} 

Understanding this present theoretical and experimental situation in
$\phi$ photoproduction, Ozaki et al.~\cite{ozaki2009} proposed a
coupled-channel effects based on the $K$-matrix formalism. They
considered the $\gamma N\to K\Lambda^*(1520)$ and $K\Lambda^*\to\phi
N$ kernels~\cite{Nam:2005uq} in the coupled-channel formalism in
addition to $\gamma N\to \phi N$ and $\phi N\to \phi N$. It is a very
plausible idea, since the threshold energy for the $K\Lambda^*$ is
quite close to that for the bump-like structure ($E_\gamma \approx
2.3$ GeV), the $\Lambda^*(1520)$ resonance may influence $\phi$
photoproduction. Moreover, the $\gamma p\to K\Lambda^*(1520)$ reaction
can be regarded as a subreaction for the $\gamma p \to K\overline{K}
p$ process together with the $\gamma p \to \phi p$ one in 
Ref.~\cite{Nam:2005uq}.
In addition, a possible nucleon resonance
($J^P=1/2^-$) with large $s\bar{s}$ content was also 
taken into account. Interestingly, the coupled-channel effects were
shown to be not enough to explain the bump-like structure $E_\gamma
\approx 2.3$ GeV. On the other hand, the bump-like structure 
was described by their possible $N^*$ resonance and was 
interpreted as a destructive interference arising from the $N^*$
resonance~\cite{Kiswandhi:2011cq,Kiswandhi:2010ub}.     

In the present work, we want to scrutinize in detail the
nontrivial hadronic contributions arising from hadronic box diagrams
in addition to Pomeron and pseudoscalar meson exchanges. Extending the
idea of Ref.~\cite{ozaki2009}, we consider seven possible box diagrams
with intermdiate $\rho N$, $\omega N$, $\sigma N$, $\pi N$, 
$K\Lambda(1116)$, $K^*\Lambda(1116)$, and $K\Lambda(1520)$ states.       
However, it is quite complicated to compute these box diagrams
explicitly, so that we use the Landau-Cutkosky
rule~\cite{Landau:1959fi,Cutkosky:1960sp}, which yields the imaginary
part of the box diagrams by its discontinuity across the branch
cut. Though their real part may contribute to the transition
amplitude, we will show that the imaginary part already illuminates
the coupled-channel effects on the production mechanism of $\gamma
p\to \phi p$ near the threshold.  The parameters such as the coupling
constants and cut-off masses of the form factors will be fixed by
describing the corresponding processes and by using experimental and
empirical data. Yet unknown parameters are varied as compared to the
present experimental data. In addition, we tune the strength of the
Pomeron amplitude near the threshold region, where the hadronic
contribution seems more significant. It is a legitimate procedure,
since the Pomeron gets more important as the energy increases. Thus,
we determine the threshold parameter in such a way that the Pomeron  
exchange becomes effective in the higher energy region. 
We did not consider any $N^*$ resonance, since we do not have much 
information on them above the $\phi N$ threshold~\cite{PDG2012}.
We will show that the coupled-channel effects are indeed essential in
explaining the recent LEPS data, which is the different conclusion
from Ref.~\cite{ozaki2009}.

The present paper is organized as follows. In Section II, we explain
the basic formalism. We show how to compute the box diagrams mentioned
above. In Section III, we present the numerical results such as the
energy dependence of the forward cross sections, the angular
distributions, and the spin observables. We also discuss how the
$K\Lambda^*(1520)$ channel can explain the bump-like structure
together with the Pomeron exchange tuned. We discuss in detail 
the spin-density matrix elements for $\phi$-photoproduction. The final 
Section is devoted to summary and outlook.  In the Appendix, we
present the definition of the spin-density matrix elements for
reference. 
\section{General Formalism}
In the present work, we will employ the effective Lagrangian approach
in addition to the Pomeron-exchange.
\begin{figure}[ht]
\centering
            \includegraphics[width=4.5cm]{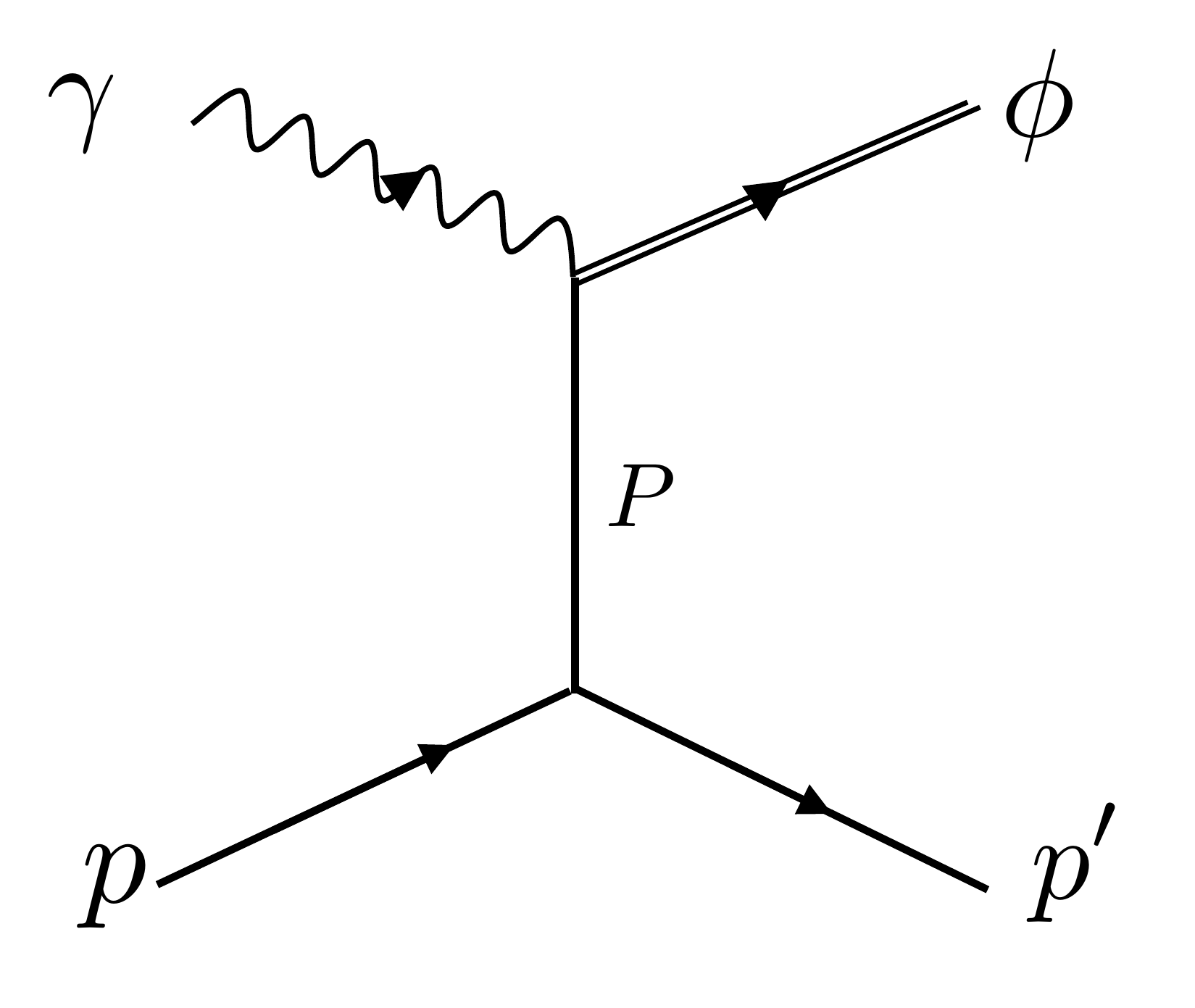} \;
            \includegraphics[width=4.5cm]{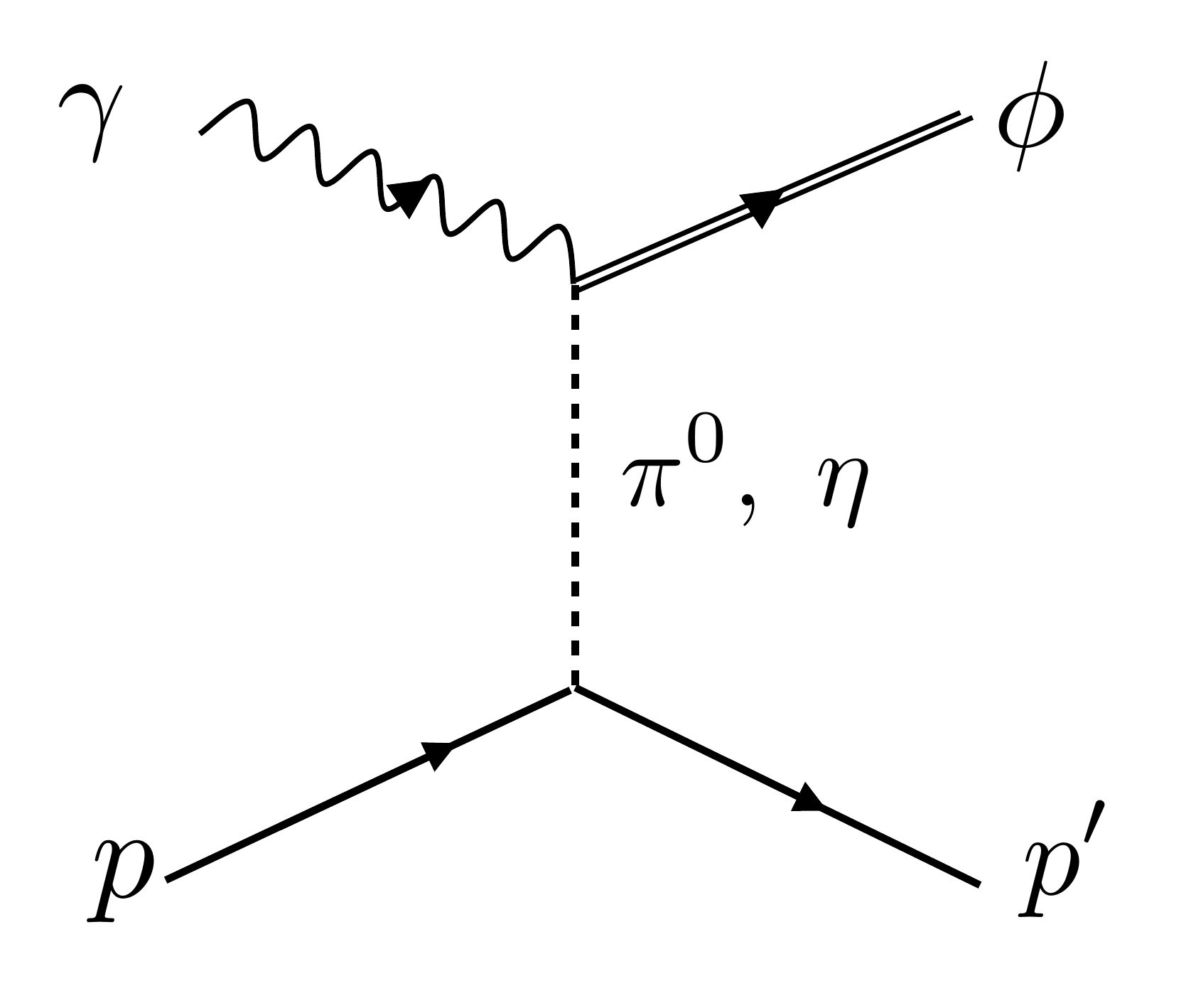}\;\;
            \includegraphics[width=5cm]{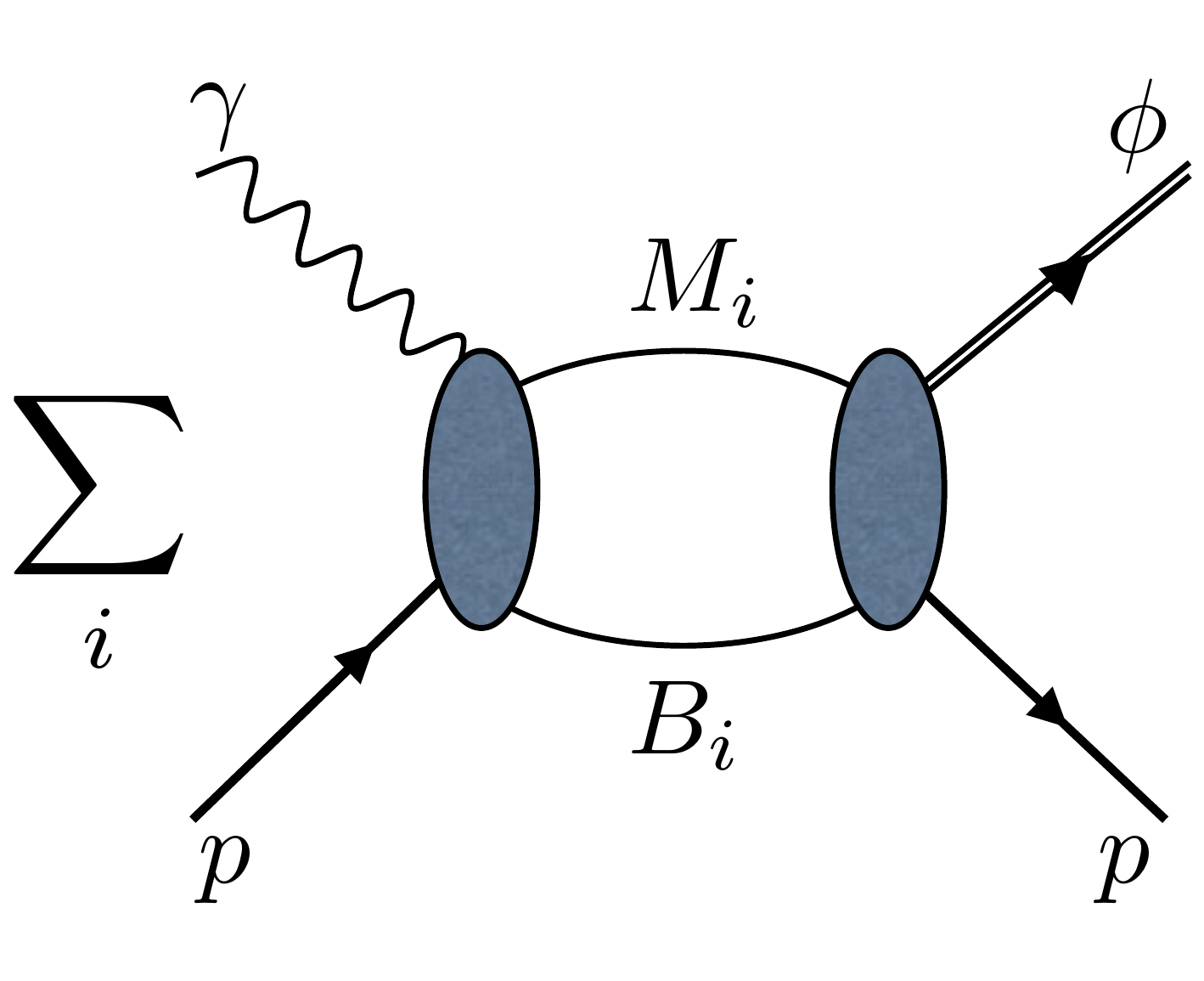}

\caption{(Color online) Relevant Feynman diagrams for $\phi$
  photoproduction: We draw, from the left, the diffractive Pomeron
  exchange, the pseudoscalar meson-exchanges, and the generic box
  diagram that includes intermediate meson $M_i$ and baryon $B_i$
  states.} 
\label{fig:1}    
\end{figure}
In Fig.~\ref{fig:1}, we draw the relevant Feynman diagrams which will be
involved in describing $\phi$ photoproduction. The first diagram
corresponds to the Pomeron-exchange, and the second one depicts
$\pi^0$- and $\eta$-exchanges. The last diagram represents generically
all the contributions from various box diagrams with intermediate
hadron states, i.e. $\rho N$,
$\omega N$, $\sigma N$, $\pi N$, $K\Lambda(1116)$, $K^*\Lambda(1116)$,  
and $K\Lambda(1520)$, among which the last one was already
considered in Ref.~\cite{ozaki2009}. From now on, we will simply
define the $\rho N$ box diagram as that with intermediate $\rho$ and
$N$ states, and so on. We also define the 4-momenta of 
the incoming photon, outgoing $\phi$, the initial (target) proton and
the final (recoil) proton as $k_1$ and $k_2$, $p_1$ and $p_2$,
respectively. In the center of mass (CM) frame, these variables are
written as $k_1=(k,\bm k)$, $k_2=(E_{\phi},\bm p)$, $p_1=(E_{p},-\bm
k)$ and $p_2=(E_{p'},-\bm p)$, where $k=|\bm k|$,
$E_\phi=\sqrt{m_\phi^2+|\bm p|^2}$, $E_p=\sqrt{m_p^2+|\bm
  k|^2}$, and $E_{p'} =\sqrt{m_{p'}^2+|\bm p|^2}$, respectively. 

\subsection{Pomeron exchange}
The amplitude of the
Pomeron-exchange~\cite{Donnachie1987,Titov:2003bk,Titov:2007fc} is
given by  
\begin{equation}
  \label{eq:1}
  \mathcal{ M}=-\bar u (p_2) \mathcal{ M}_{\mu \nu} u(p_1)
  \epsilon_{\phi}^{*\mu} \epsilon_{\gamma}^{\nu} , 
\end{equation}
where $\epsilon_{\phi}$ and $\epsilon_{\gamma}$ are the polarization
vectors of the $\phi$ meson and photon. $\mathcal{ M}_{\mu \nu}$ is 
\begin{equation}
  \label{eq:2}
   \mathcal{ M}^{\mu \nu}=M(s,t)\Gamma^{\mu \nu},
\end{equation}
where the transition operator $\Gamma^{\mu \nu}$ is defined as 
\begin{eqnarray}
   \Gamma^{\mu \nu}&=& \rlap{/}{k}_{\gamma}
   \bigg( g^{\mu\nu}-\frac{p_{3}^{\mu} p_{3}^{\nu}}{p_{3}^2} \bigg)
   -\gamma^\nu \bigg( k_\gamma^\mu -p_{3}^\mu 
   \frac{k_{1}\cdot p_{3}}{p_{3}^2} \bigg) \cr
   && -\bigg(  p_{3}^\nu -\bar{p}^\nu 
   \frac{k_\gamma \cdot p_3}{\bar{p}\cdot k_1} \bigg)
   \bigg( \gamma^\mu -\frac{\rlap{/}{p}_3 p_{3}^{\mu}}{p_{3}^2}
   \bigg),    \label{eq:3}
\end{eqnarray}
with $\bar{p}=(p_1+p_2)/2$. Note that the Pomeron amplitude preserves 
gauge invariance  $k_1^\nu \mathcal{ M}_{\mu\nu}=0$. 
The corresponding invariant amplitude $M(s,t)$ in Eq.(\ref{eq:2}) is
written as 
\begin{equation}
  \label{eq:4}
  M(s,t)=C_p F_N(t)F_{\phi}(t)\frac{1}{s}\Big(
  \frac{s-s_{\mathrm{th}}}{4} \Big)^{\alpha_p(t)} 
  \mathrm{exp}\Big( -\frac{i\pi}{2}\alpha_p(t) \Big),    
\end{equation}
where $s=(k_1 +p_1)^2$ and $t=(k_1-k_2)^2$. 
$F_N(t)$ is the isoscalar form factor of the nucleon, whereas
$F_{\phi}(t)$ is the form factor for the photon-$\phi$ meson-Pomeron
vertex. They are parameterized, respectively, as 
\begin{eqnarray}
  F_N(t)&=&\frac{4M_N^2 -a    _N^2 t}{(4M_N^2-t)(1-t/t_0)^2}, \cr
  F_{\phi}(t)&=&\frac{2 \mu_0^2}{(1-t/M_\phi^2)(2\mu_0^2 +M_{\phi}^2
    -t)}. 
  \label{eq:5}
\end{eqnarray}
The Pomeron trajectory $\alpha_p(p)=1.08+0.25\, t$ in Eq.(\ref{eq:4}) 
is determined from hadron elastic scattering in the high-energy
region. The prefactor $C_p$ in Eq.(\ref{eq:4}) governs the overall
strength of the amplitude and $s_{\mathrm{th}}$ determines 
the starting energy at which the Pomeron-exchange comes into play.  We
will discuss the determination of these two parameters in
Section~\ref{sec:3}. 
\subsection{$\pi$- and $\eta$-exchanges} 
To calculate pseudoscalar meson $(\varphi=\pi^0,~\eta)$ exchange in
the $t$ channel, we introduce the following effective Lagrangians:
\begin{eqnarray}
  \mathcal{ L}_{\phi \gamma \varphi}&=&\frac{e}{m_\phi}g_{\phi \gamma
    \varphi}   \epsilon^{\mu \nu \alpha \beta} \partial_\mu
  \phi_\nu \partial_\alpha A_\beta  \varphi~,  \cr
  \mathcal{ L}_{\varphi NN}&=&\frac{g_{\varphi NN}}{2M_N}\bar N
  \gamma_\mu \gamma_5  N \partial^\mu \varphi ,   \label{eq:6}
\end{eqnarray}
where $\phi_\nu$, $A_\beta$, and $N$ denote the $\phi$ vector meson,
photon, and nucleon fields, respectively. $m_\phi$ and $M_N$ stand for
the $\phi$ meson and nucleon masses respectively. $e$ represents the
electric charge. The $t$-channel amplitude then takes the following
form: 
\begin{equation}
  \label{eq:7}
  \mathcal{ M}= \frac{eg_{\varphi  
    NN}g_{\phi \gamma \varphi}}{m_\phi} \frac{i F_{\varphi NN}(t)
    F_{\phi\gamma\varphi}}{t-M_{\varphi}^2}
  \bar u(p_2) (\rlap{/}{k}_1-\rlap{/}{k}_2)\gamma_5 u(p_1) 
  \epsilon^{\mu\nu\alpha\beta}k_{2\mu} \epsilon_{\phi \nu}^* k_{1\alpha}
     \epsilon_{\gamma \beta},   
\end{equation}
where $r$ is the four momentum of an exchanged pseudoscalar meson.
We introduce the monopole-type form factors for each vertex
$F_{\varphi NN}(t)$ and $F_{\phi\gamma\varphi}$ defined as 
\begin{equation}
F_{\varphi NN} (t) \;=\; \frac{\Lambda_{\varphi NN}^2 -
  M_\varphi^2}{\Lambda_{\varphi NN}^2 - t}, \;\;\;
F_{\phi \gamma \varphi} (t) \;=\; \frac{\Lambda_{\phi \gamma
    \varphi}^2 -  M_\varphi^2}{\Lambda_{\phi \gamma \varphi}^2 - t}.   
\end{equation}

The coupling constants and the cutoff masses for the
pseudoscalar-exchange, we follow Ref.~\cite{Titov:1999eu}: $g_{\pi
  NN}=13.26$, $g_{\eta NN}=3.527$ for the 
$\pi NN$ and $\eta NN$ coupling constants, respectively.  The cutoff
masses are taken to be 
$\Lambda_{\pi NN}=0.7\,\mathrm{GeV}$ and $\Lambda_{\eta
  NN}=1\,\mathrm{GeV}$.
Though these values are somewhat different from the
phenomenological nucleon-nucleon
potentials~\cite{Machleidt:1989tm,Rijken:2006en}, the effects 
of the pseudoscalar meson-exchanges on $\phi$ photoproduction are rather
small. Thus, we will take the values given above typically used in
$\phi$ photoproduction. 
Those of the coupling constants for the $\phi\gamma\varphi$ 
vertices are determined by using the radiative decays of the $\phi$
meson to $\pi$ and $\eta$. Using the data from the Particle Data Group
(PDG)~\cite{PDG2012}, one can find $g_{\phi\gamma\pi}=-0.141$ amd
$g_{\phi\gamma\eta}= -0.707$. The negative signs of
these coupling constants were determined by the phase conventions in
SU(3) symmetry as well as by $\pi$ photoproduction~\cite{Titov:1999eu}.  
We choose the cut-off masses for the $\phi\gamma\pi$ and
$\phi\gamma\eta$ form factors as follows: $\Lambda_{\phi\gamma
  \pi}=0.77$ GeV and $\Lambda_{\phi\gamma  \eta}=0.9$ GeV,
respectively. 
\subsection{ $K^+  \Lambda(1520)$ box diagram} 
In addition to the Pomeron- and pseudoscalar meson-exchanges, we 
include the seven different box diagrams: $\rho N$, $\omega N$,
$\sigma N$, $\pi N$, $K\Lambda(1116)$, $K^*\Lambda(1116)$, and
$K\Lambda(1520)$. Since the $K\Lambda(1520)$ box diagram is the 
most significant one among several possibie box diagram in describing
$\phi$ photoproduction, we first discuss the $K^+\Lambda(1520)$ one
and then deal with all other box diagrams in the next subsection. In
The $\gamma N\to K^+ \Lambda(1520)$ process was investigated within an
effective Lagrangian method in Ref.~\cite{Nam:2005uq} of which the
results were in good agreement with the experimental data. Thus, we
will take the formalism developed in Ref.~~\cite{Nam:2005uq} so that
we may take into account the $K\Lambda(1116)$ coupled-channel effects
more realistically. 

The effective Lagrangians for $\gamma N\to K^+ \Lambda(1520)$ are
written as 
\begin{eqnarray}
  \mathcal{L}_{KN\Lambda^*}&=&  \frac{g_{KN\Lambda^*}}{M_K} 
      \bar N \gamma_5 \partial_\mu K^+ \Lambda^{* \mu},\cr
  \mathcal{L}_{\phi K N \Lambda^*}
  &=&-i\frac{g_{KN\Lambda^*}}{M_K}g_{\phi KK}  
      \bar N \gamma_5 \partial_\mu K^+ \Lambda^{* \mu},\cr
  \mathcal{L}_{\phi KK}&=&-ig_{\phi KK}(\partial^\mu K^- K^+
  -\partial^\mu K^+ K^-)\phi_\mu,\cr
  \mathcal{L}_{\phi NN}&=&-g_{\phi NN} \bar N\left[\gamma_\mu
    \phi^\mu      -\frac{\kappa_\phi}{2M_N} \sigma^{\mu
      \nu} \partial_\nu \phi_\mu     \right] N ,  \cr
  \mathcal{L}_{\gamma KK}&=&-ie(\partial^\mu K^- K^+ -\partial^\mu
  K^+ K^-)A_\mu,\cr
  \mathcal{L}_{\gamma NN}&=&-e \bar N \left[ \gamma^\mu 
     -\frac{\kappa_N }{2M_N} \sigma^{\mu \nu} \partial_\nu 
   \right]A_\mu N,\cr
  \mathcal{L}_{\gamma K N \Lambda^*}&=& -i\frac{eg_{KN\Lambda^*}}{M_K} 
      \bar N \gamma_5 \partial_\mu K^+ \Lambda^{* \mu},
  \label{eq:9}
\end{eqnarray}
where $K$ and $\Lambda^{*\mu}$ denote the $K$ meson and
$\Lambda(1520)$ fields. For $\Lambda(1520)$, we utilize the
Rarita-Schwinger formalism. $M_K$ is the kaon mass. The $KN\Lambda^*$ 
coupling constant is taken from Ref.~\cite{Nam:2005uq}, since we use
the amplitude derived in it. The $\phi KK$ coupling constant can be
determined from the experimental data for the decay width
$\Gamma_{\phi\to KK}$. On the other hand, $g_{\phi NN}$ is not
much known experimentally. Recent experiments measuring the strange
vector form factors imply that the strange quark gives almost no
contribution to the nucleon electromagnetic (EM) form
factors~\cite{Ahmed:2011vp}. One can deduce from this experimental
fact that the $\phi NN$ coupling constant should be very
small. In Ref.~\cite{Meissner:1997qt}, the $\phi NN$ was estimated by
using a microscopic hadronic model with $\pi\rho$ continuum: $g_{\phi
  NN}=\pm 0.25$ and $\kappa_\phi=0.2$, which are compatible with the
recent data for the strange vector form factors. Thus, we will take
these values in the present work. However, note that the $s$-channel
contribution with the $\phi NN$ vertex is almost negligible. In
Table~\ref{tab:1}, the relevant strong coupling constants and
anomalous magnetic moments are listed.    
\begin{table}[ht]
  \begin{center}
    \caption{The strong coupling constants and anomalous magnetic
      moments used in the present work.} 
    \label{tab:1}
    \begin{tabular}{lll}
\hline \hline
$g_{KN\Lambda^*}$ {}\hspace{1cm}  \hspace{1cm} {}
& 11{}\hspace{1cm}\hspace{1cm} {}
&Ref.~\cite{Nam:2005uq} \\
$g_{\phi KK}$ &4.7& Ref.~\cite{PDG2012} \\
$g_{\phi NN}$ &0.25& Ref.~\cite{Meissner:1997qt} \\
\hline
$\kappa_{p}$&1.79&Ref.~\cite{PDG2012}\\
$\kappa_{\phi}$&0.2& Ref.~\cite{Meissner:1997qt}\\
\hline \hline
    \end{tabular}
  \end{center}
\end{table}

Based on the effective Lagrangians given in Eq.(\ref{eq:9}), we can 
write down the amplitude for the $K^+ \Lambda^*(1520)$ box diagram.
It contains both real and imaginary parts. The real part is divergent,
which is also the case for other box diagrams and the rigorous
calculation is rather involved. Thus we consider that the real part
can be taken into account effectively by the reenormalization of
various coupling constants, and calculate only the imaginary part
explicitly. The reasoning behind is similar to the concept of K-matrix
formalism for the S-matrix. Physically, the imaginary part corresponds
to rescattering and is obtained by the Landau-Cutkosky rule, Ref.
\cite{Landau:1959fi, Cutkosky:1960sp}.

\begin{figure}[ht]
\centerline{
       \includegraphics[width=6cm]{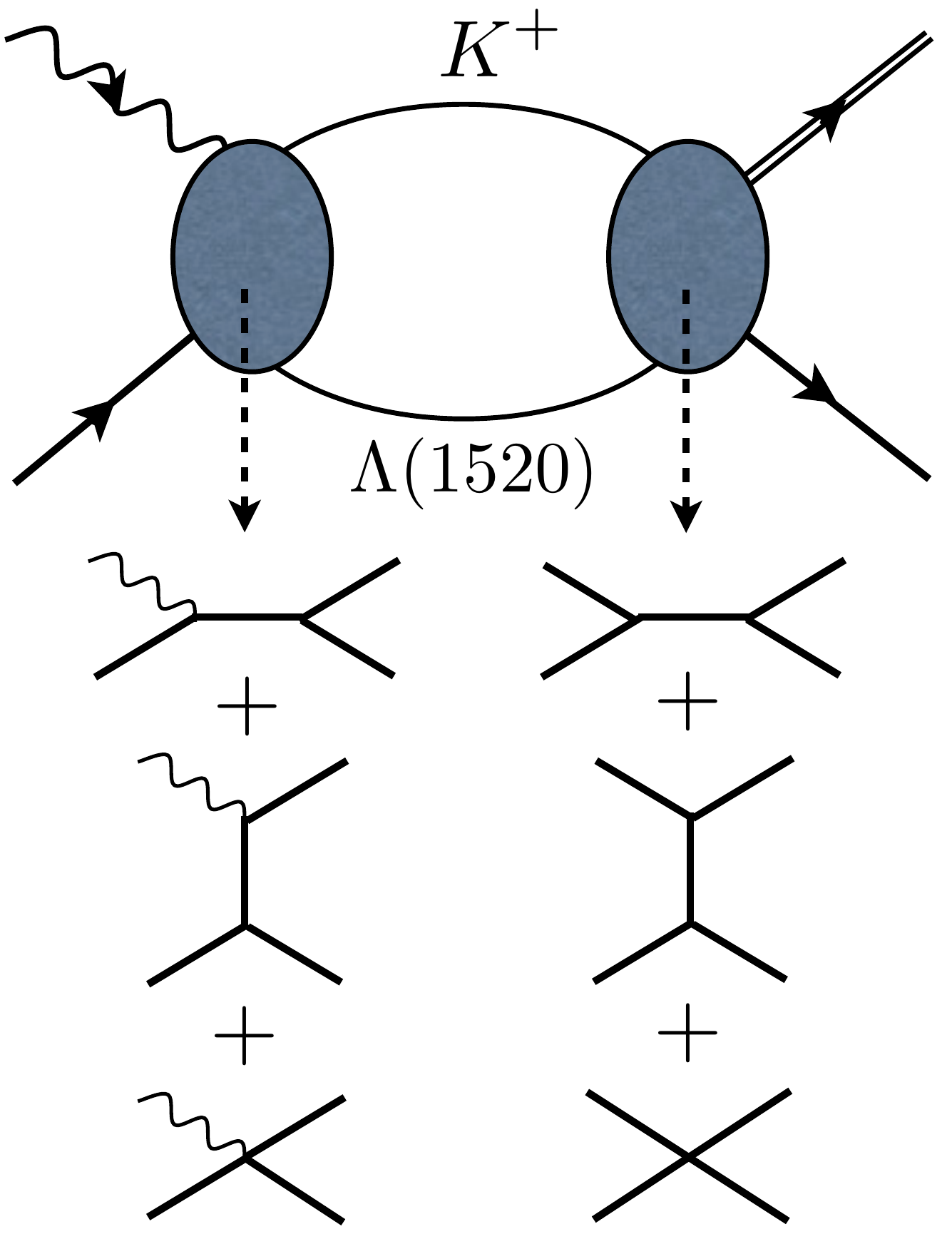}
}
    \caption{Feynman diagrams for the $K^+ \Lambda(1520)$ box. The
      form factors are introduced in a gauge-invariant way.}
  \label{fig:2}      
\end{figure}

\begin{figure}[ht]
  \begin{center}       
       \includegraphics[width=8cm]{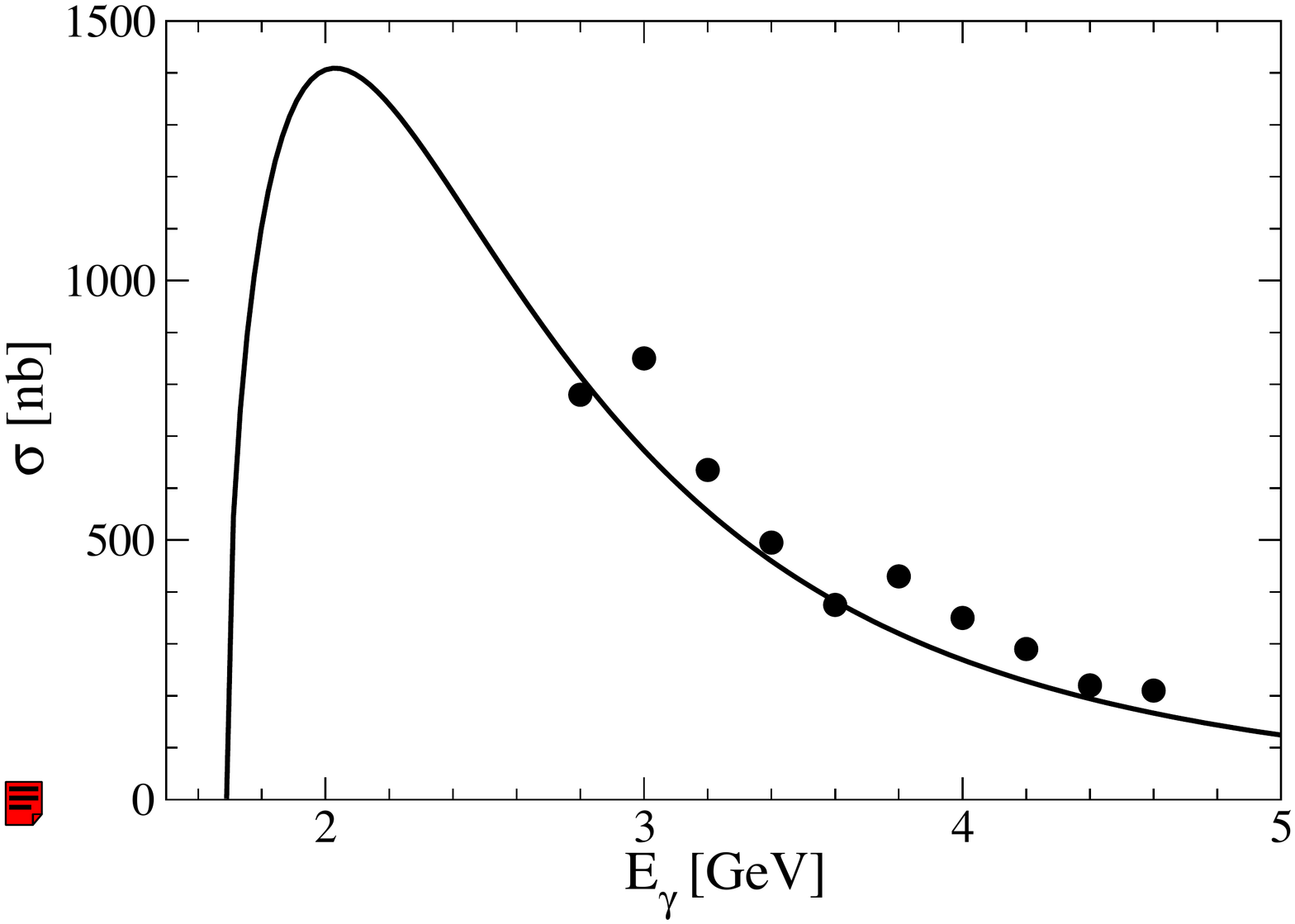}
  \end{center}
  \vspace{-20pt}
  \caption{Total cross-section of the $\gamma p \to K \Lambda(1520)$ 
  reaction as compared to the
  experimental data \cite{Adelseck1985}.}
\label{fig:3}
\end{figure}

Having computed the Lorentz-invariant phase space volume
factors, we obtain the imaginary part of the amplitude as  
\begin{equation}
  \label{eq:10}
\mathrm{Im}\mathcal{ M}_{K^+ \Lambda^* \mathrm{ box}}
\;=\; -\frac{1}{8\pi}\frac{r}{\sqrt{s}} \int \frac{d \Omega}{4 \pi}
\mathcal{ M}_L(\gamma p \to K^+ \Lambda^*) \mathcal{ M}_R^\dagger (K^+ 
\Lambda^* \to \phi p),
\end{equation}
where $r$ is the magnitude of the $K^+$ momentum. This imaginary part
of the amplitude is schematically drawn in Fig.~\ref{fig:2}.
The shaded ellipse in the left-hand side represents the invariant
amplitude for $\gamma p \to K^+ \Lambda^*$, which is basically the
same as that of Ref.~\cite{Nam:2005uq} except for different form
factors as will be explained later. It consists of three different
types of the Feynman diagrams as shown below the left dashed arrow. 
On the other hand, the right ellipse stands for the $K^+ \Lambda^* \to
\phi p$ process that contains the diagrams below the right arrow,
generically. 
Note that we use a similar method as in Ref.~\cite{ozaki2009} but we
choose the different form factors and parameters. The corresponding
invariant amplitudes $\mathcal{ 
  M}_L(\gamma p \to K^+ \Lambda^*)$ and $\mathcal{ M}_R (K^+ \Lambda^*
\to \phi p)$ with the form factors are defined as follows: 
\begin{eqnarray}
\mathcal{ M}_L (\gamma p \to K^+ \Lambda^*) &=& (\mathcal{ M}_{L,s}+
\mathcal{ M}_{L,t} 
   + \mathcal{ M}_{L,c} )F_L(s,t)  , \cr
\mathcal{ M}_R (K^+ \Lambda^* \to \phi p)&=& (\mathcal{ M}_{R,s}+
\mathcal{ M}_{R,t}    + \mathcal{ M}_{R,c})F_R(s,t),
\label{eq:13}
\end{eqnarray}
where $\mathcal{M}_{L,s}$ ($\mathcal{M}_{R,s}$), $\mathcal{ M}_{L,t}$
($\mathcal{ M}_{R,t}$), and $\mathcal{M}_{L,c}$ ($\mathcal{M}_{R,c}$)
represent the $s$-channel, the $t$-channel, and the  
contact-term contributions to the $\gamma p \to K^+ \Lambda^*$
($K^+\Lambda^*\to \phi p$) process, respectively: 
\begin{eqnarray}
  \label{eq:14}
\mathcal{ M}_{L,s} &=&  \frac{eg_{KN\Lambda^*}}{M_K} \bar u^\mu 
k_{2\mu} \gamma_5
\frac{\rlap{/}{k}_1+\rlap{/}{q}+M_N}{q^2-M_N^2} \rlap{/}{\epsilon}u(p_1),     
  \cr
  &&  +\,\frac{e\kappa_p g_{KN\Lambda^*}}{2M_N M_K}
  \bar u^\mu k_{2\mu} \gamma_5 \frac{\rlap{/}{q}+M_N}{q^2-M_p^2}
  \rlap{/}{\epsilon}\rlap{/}{k}_1 u(p_1), \cr
\mathcal{ M}_{L,t}&=& -\frac{2eg_{KN\Lambda^*}}{M_K} \bar u^\mu \gamma_5
u(p_1) \frac{q_{K}^{\mu}}{t_{K}-M_K^2},\cr
\mathcal{ M}_{L,c} &=&   \frac{eg_{KN\Lambda^*}}{M_K}\bar u^\mu \epsilon_\mu 
\gamma_5 u(p_1),\cr
\mathcal{ M}_{R,s} &=&  -i\frac{g_{KN\Lambda^*}g_{\phi NN}}{M_K}\bar u(p_2)
\rlap{/}{\epsilon}_{\phi}^{*} \frac{\rlap{/}{q}+M_p}{q^2-M_p^2}\gamma_5 k_{1}^{\alpha}
u^{\alpha}(p_1),\cr
&&+\,i\frac{g_{KN\Lambda^*}g_{\phi NN}}{M_K}\frac{\kappa_{\phi}}{2M_p}
\bar u(p_2) \rlap{/}{k}_2 \rlap{/}{\epsilon}_{\phi}^{*}
\frac{\rlap{/}{q}+M_p}{q^2-M_p^2}\gamma_5 k_{1}^\alpha u^\alpha (p_1),\cr
\mathcal{ M}_{R,t}  &=&  \frac{-ig_{KN\Lambda^*}g_{\phi KK}}{M_K}   
  \frac{2k_1 \cdot \epsilon_{\phi}^*}{q_K^2 -M_K^2}
  \bar u (p_2) \gamma_5 q_t^\alpha u^\alpha (p_1),\cr
\mathcal{ M}_{R,c} &=& \frac{-ig_{KN\Lambda^*}g_{K NN}}{M_K}
\bar u(p_2) \gamma_5 \epsilon_{\phi}^{*\mu}u^\mu (p_1).  
\end{eqnarray}
We introduce the form factors $F_R(s,t)$ and $F_L(s,t)$ for 
 $\mathcal{ M}_R$ and $\mathcal{ M}_L$, respectively, in particular,
 in a gauge-invariant manner for the $\gamma p\to K^+\Lambda^*$
 rescattering: 
 \begin{eqnarray}
   \label{eq:11}
F_R(s,t)&=&
\left[ \frac{n_{1}\Lambda_{1}^4}{n_{1}\Lambda_{1}^4+(s-M_p^2)^2}
\right]^{n_{1}} 
\left[ \frac{n_{2}\Lambda_{2}^4}{n_{2}\Lambda_{2}^4+t^2}
\right]^{n_{2}},\cr
F_L(s,t)&=&
\left[ \frac{n_{3}\Lambda_{3}^4}{n_{3}\Lambda_{3}^4+(s-M_p^2)^2}
\right]^{n_{3}} \left[ \frac{n_{4}\Lambda_{4}^4}{n_{4}\Lambda_4^4+ t^2}
\right]^{n_{4}} ,
 \end{eqnarray}
where the cut-off masses $\Lambda_i$ and powers $n_i$ are fitted to
the experimental data for the $\gamma p\to K^+\Lambda^*$ and $\gamma p 
\to \phi p$,  which are listed in Table~\ref{tab:2}. 
\begin{table}[ht]
  \begin{center}
    \caption{Cut-off parameters used in Eq.(\ref{eq:11})}
    \label{tab:2}
    \begin{tabular}{ll}
\hline \hline
$n_{1}${}\hspace{1cm}  \hspace{1cm} {} & 1  \\
$n_{2}$&1   \\
$n_{3}$ &2   \\
$n_{4}$ &1   \\
$\Lambda_{1}$&0.8 GeV \\
$\Lambda_{2}$&0.8 GeV \\
$\Lambda_{3}$ &1.0 GeV  \\
$\Lambda_{4}$ &1.0 GeV  \\
\hline \hline
    \end{tabular}
  \end{center}
\end{table}
In Fig.~\ref{fig:3}, we draw the numerical result of the total cross
section for $\gamma p\to K^+ \Lambda^*$ in comparison with the
experimental data taken from Ref.~\cite{Adelseck1985}. It is in good
agreement with the data. 
\subsection{All other box diagrams}
\begin{figure}[ht]
  \centering
     \includegraphics[width=4.25cm]{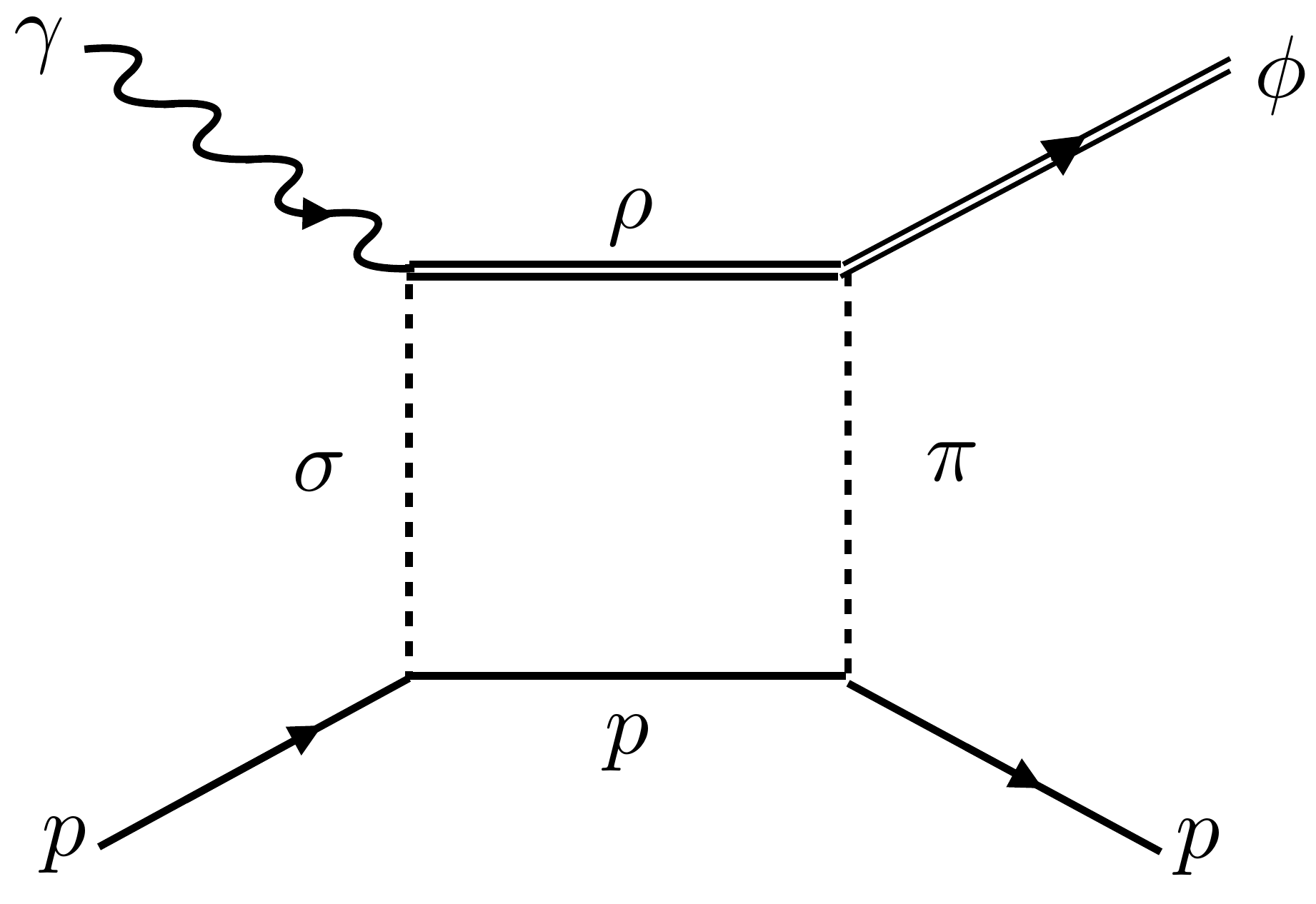}\;\;\;
     \includegraphics[width=4.25cm]{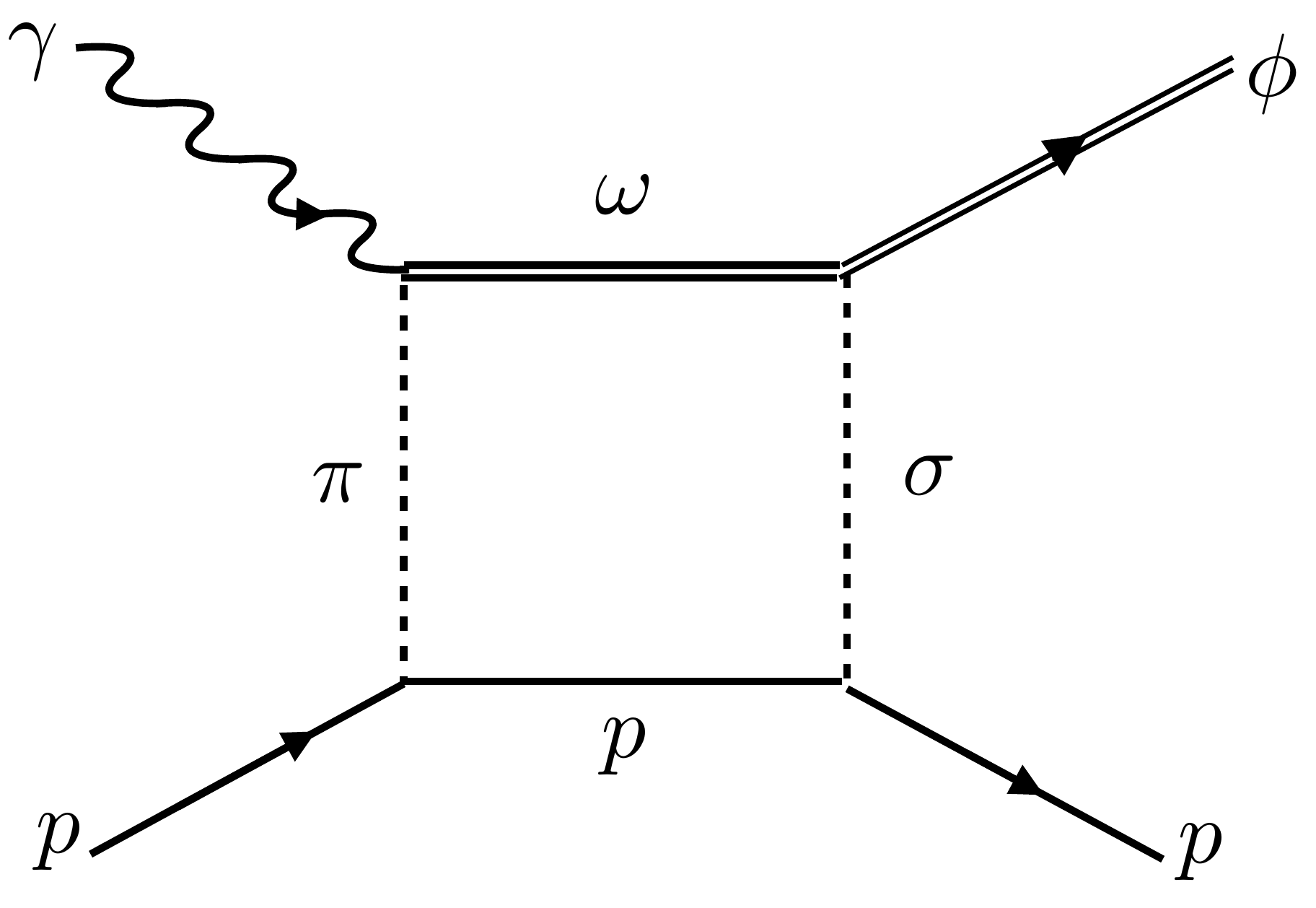}\;\;\;
     \includegraphics[width=4.25cm]{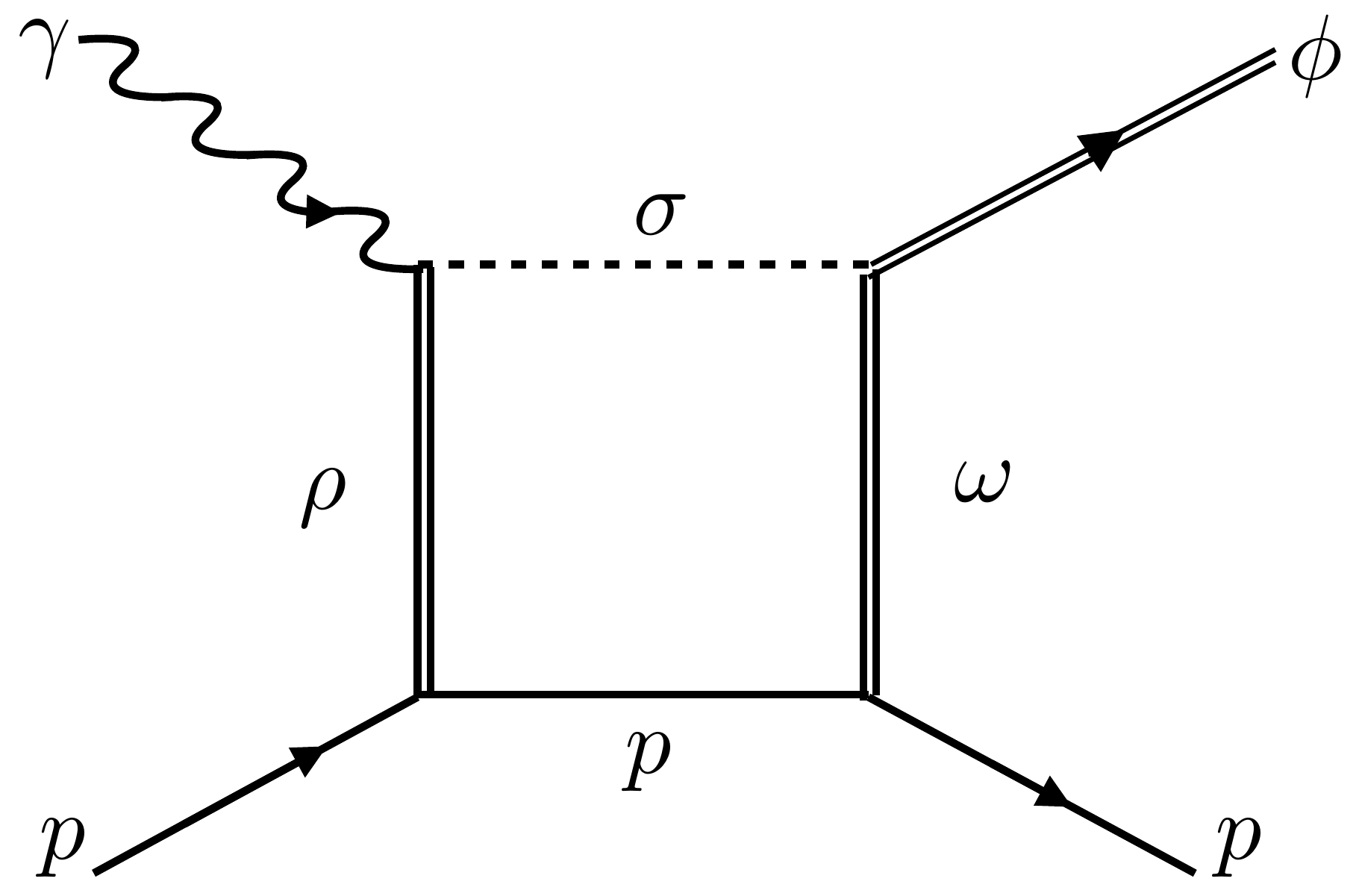}\\ \vspace{0.5cm}
     \includegraphics[width=4.25cm]{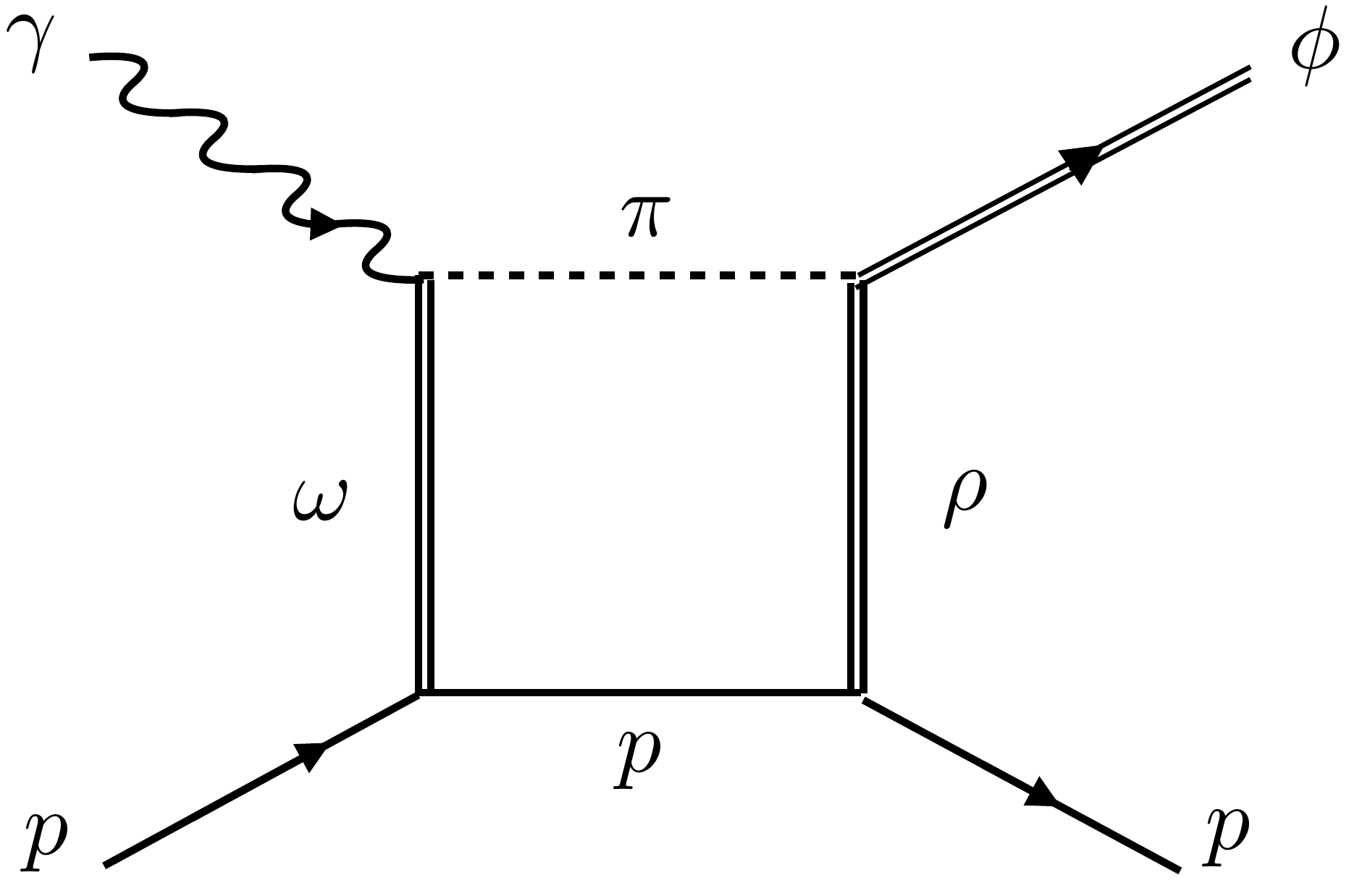}\;\;\;
     \includegraphics[width=4.25cm]{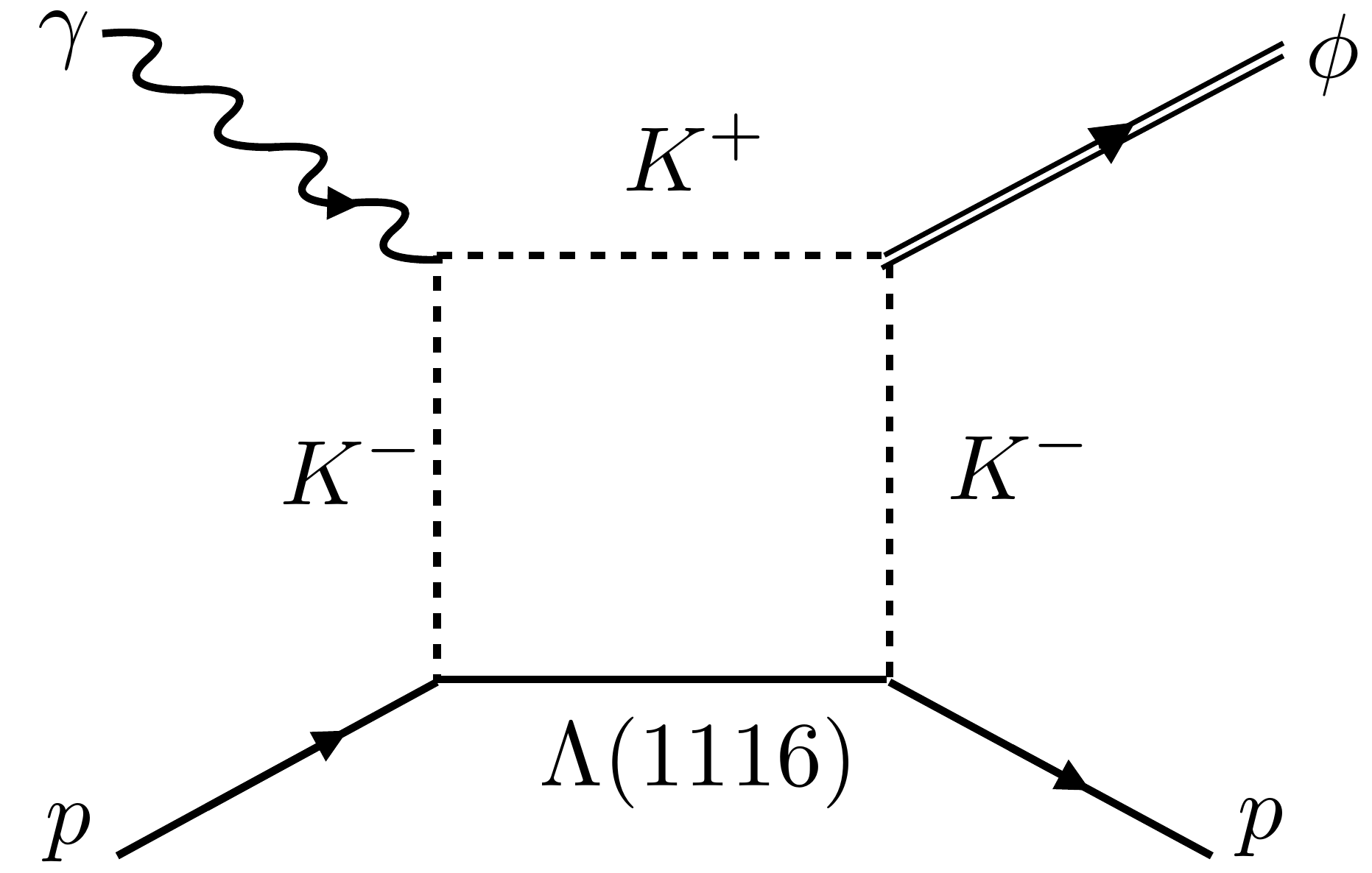}\;\;\;
     \includegraphics[width=4.25cm]{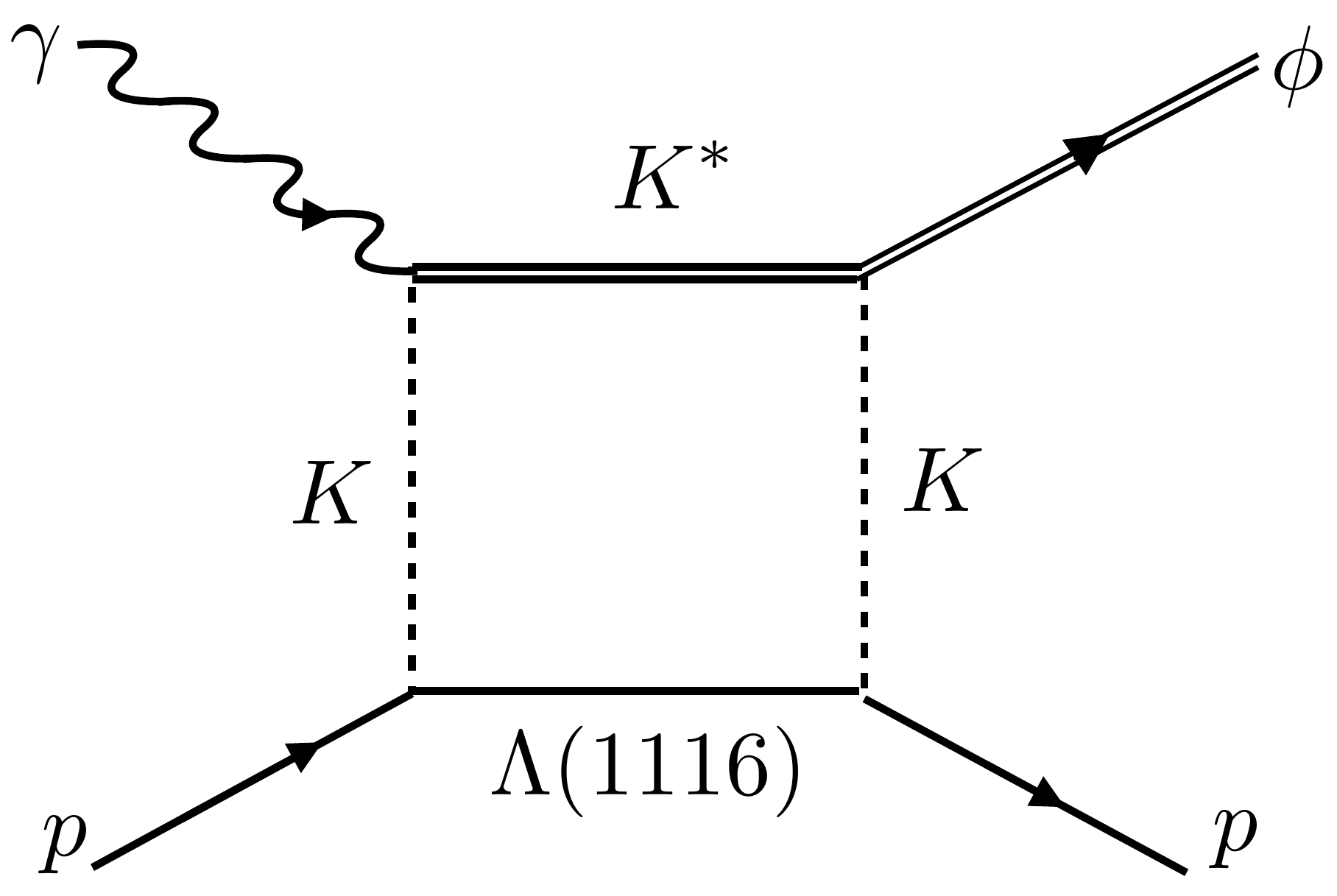}
\caption{Feynman diagrams for the six hadronic box contributions.}
\label{fig:4}
\end{figure}

\begin{figure}[ht]
  \begin{center}       
       \includegraphics[width=8cm]{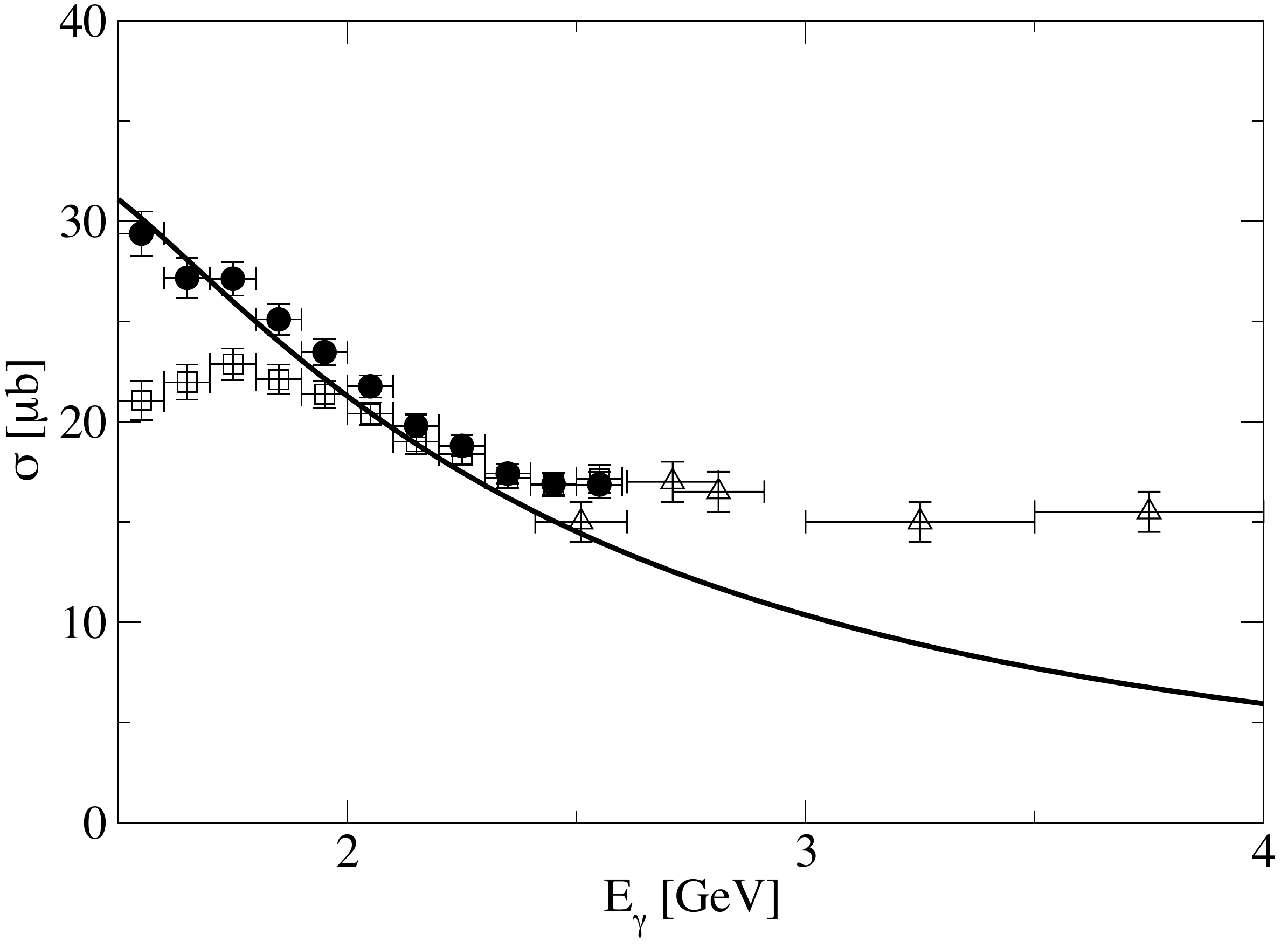}
  \end{center}
  \caption{Total cross-section of the $\gamma p \to \rho^0 p$ reaction.
   The solid curve depicts the present result obtained from the
   $t$-channel $\sigma$-exchange diagram. 
 The closed circles and the open squares are taken from
 Ref.~\cite{Wu:2005wf}, where as the open triangles represent those
 from Ref.~\cite{ABBHHM:1968aa}.
}
\label{fig:5}
\end{figure}

In the same manner as done for the $K^+\Lambda^*$ box diagram, we
consider the six intermediate box diagrams as shown in
Fig.\ref{fig:4}, i.e.  the $\rho N$, $\omega N$, $\sigma N$, $\pi N$,
$K\Lambda(1116)$, and $K^*\Lambda(1116)$ box diagrams.  
$\rho$ photoproduction has been studied 
theoretically~\cite{Friman:1995qm,Kaneko:2011bd,
  Kiswandhi:2011ei,Kiswandhi:2010ub} in which 
the contributions of the $t$-channel $\pi$- and $\sigma$-exchanges
were considered and $\sigma$-exchange was found to be the dominant
one, since it selects the isovector part of the EM current. Thus, we
take into account the $\rho p$ box diagram with the $\sigma$-  
and $\pi$-exchanges in the $t$-channel, as shown in the first diagram
of Fig.~\ref{fig:4}. We will show later in Fig.~\ref{fig:5} that
indeed the $\sigma$-exchange describes qualitatively well the $\gamma
p \to \rho p$ reaction. In Ref.~\cite{Friman:1995qm} $\omega$
photoproduction was also discussed within the same framework. In
contrast to the $\gamma p\to \rho p$ reaction, the $\pi$-exchange
appeared to be dominant, since it picks up the isoscalar part of the
EM current. Correspondingly, we consider the $\omega p$ box
contribution as in the second diagram of Fig.~\ref{fig:4},
where $\omega$ is produced by the one pion exchange.  
The $\sigma p$ and $\pi p$ box diagrams are 
obtained by reversing the $\rho p$ and $\omega p$ box diagrams. 
The $\gamma p\to K \Lambda(1116)$ and $\gamma p\to K^*\Lambda(1116)$
reactions were measured by several experimental
collaborations~\cite{Glander:2003jw,Bradford:2005pt,Sumihama:2005er,
Hicks:2007zz,Achenbach:2011rf,Hicks:2010pg} and were investigated
theoretically~\cite{Janssen:2001pe,Oh:2006hm,Oh:2006in,
Yu:2011fv,Kim:2011rm}. 
While we consider all the relevant diagrams
for the $K \Lambda^*(1520)$ box contribution because of its
significance, we will take into account only the $K$-exchange diagrams
in the $t$-channel for the $K\Lambda$ and $K^*\Lambda$ box diagrams,
since these two box diagrams turn out to have tiny effects on $\phi$
photoproduction.    

\begin{figure}[ht]
  \begin{center}       
       \includegraphics[width=8cm]{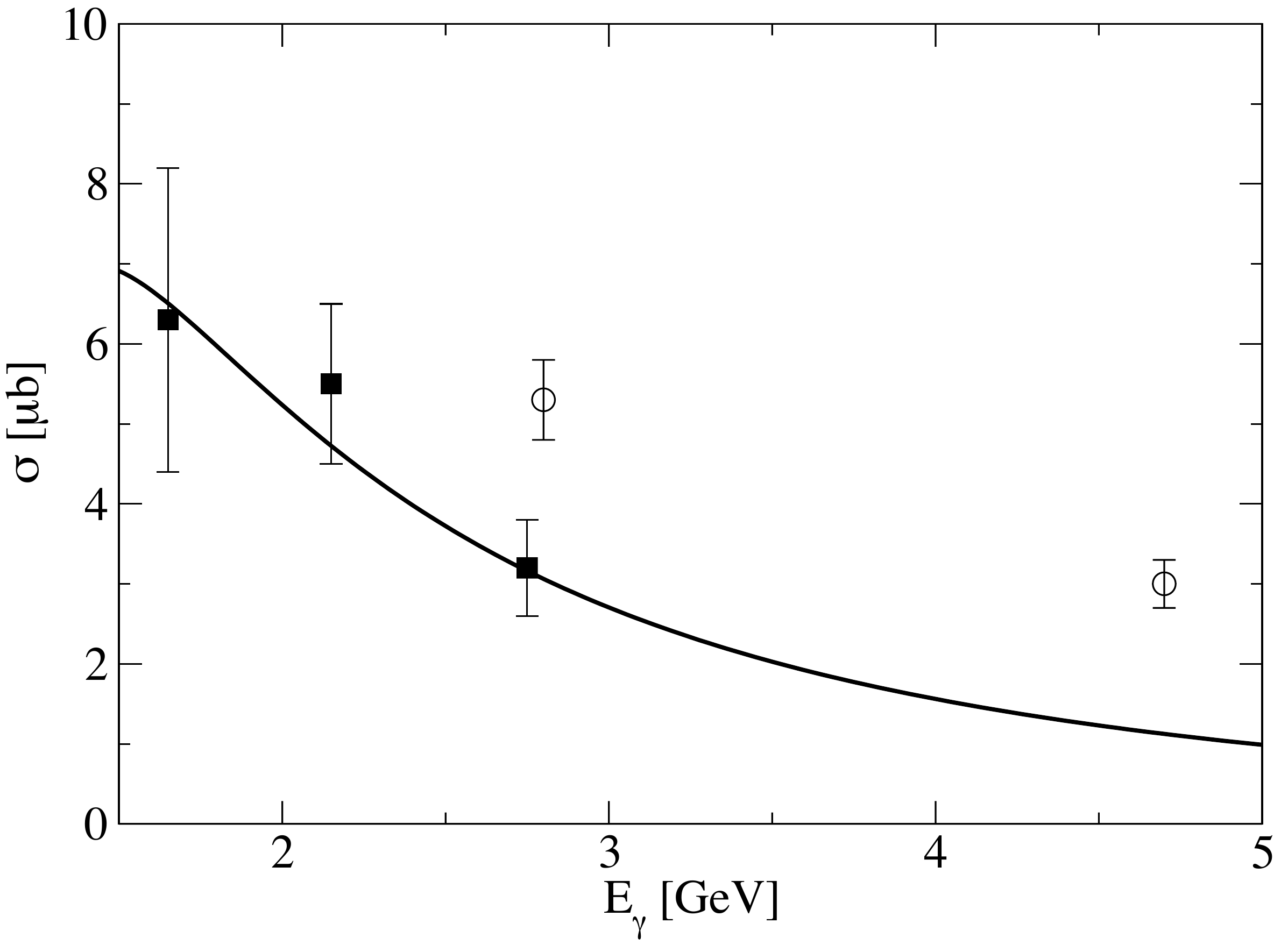}
  \end{center}
  \caption{Total cross-section of the $\gamma p \to \omega p$
    reaction. The solid curve depicts the 
present result obtained from the $t$-channel $\pi$-exchange
diagram. The closed squares denote the experimental data from
Ref.~\cite{Ballam1973} whereas the open circles represent those from
Ref.~\cite{BHMITPWBubbleChamberGroup:1967zz}.   
}
\label{fig:6}
\end{figure}

The relevant effective Lagrangians for these box diagrams are given as 
follows: 
\begin{eqnarray}
  \label{eq:8}
\mathcal{ L}_{\gamma \rho \sigma}&=& \frac{g_{\gamma \rho
    \sigma}}{m_{\rho}} 
  [\partial_{\mu}A_{\nu}\partial^{\mu} \rho^{\nu}-
  \partial_{\mu}A_{\nu}\partial^{\nu} \rho^\mu ] \sigma, \cr
\mathcal{ L}_{\sigma NN}&=& g_{\sigma NN}\bar N N \sigma,  \cr
\mathcal{ L}_{\pi^0 NN}&=&-ig_{\pi NN} \bar N \gamma_5 \tau_3 N
\pi^0, \cr
\mathcal{ L}_{\pi \rho \phi}&=& \frac{g_{\pi \rho \phi}}{m_\phi}
\epsilon_{\mu \nu \alpha \beta} 
\partial^{\nu} \phi^\mu \partial^\beta \rho^\alpha
\pi_0,\cr
\mathcal{ L}_{\omega \phi \sigma}&=& \frac{g_{\omega \phi
    \sigma}}{m_{\phi}}
(\partial_{\mu}\omega_\nu \partial^{\mu}\phi^{\nu} 
-\partial_{\mu}\omega_\nu \partial^{\nu}\phi^{\mu}),\cr
\mathcal{ L}_{\gamma \omega \pi}&=& \frac{g_{\gamma \omega \pi}}{m_{\omega}}
\epsilon_{\mu \nu\alpha \beta} \partial^\nu A^\mu \partial^\beta
\omega^\alpha \pi^0  ,\cr
\mathcal{ L}_{V NN}&=&-g_{V NN}\bar N 
\left(\gamma_\mu V^\mu -\frac{\kappa_V}{2M_N}\sigma^{\mu \nu}
\partial_{\nu} V_\mu \right) N~,\quad(V=\omega,~ \rho),\cr
\mathcal{ L}_{\gamma KK}&=&-ie \left[ (\partial^\mu K^+)K^- -
(\partial^\mu K^-)K^+ \right]A_\mu,\cr
\mathcal{ L}_{\phi KK}&=&ig_{\phi KK} \left[ (\partial^\mu K^+)K^- -
(\partial^\mu K^-)K^+ \right]\phi_\mu,\cr
\mathcal{ L}_{KN\Lambda}&=&-ig_{KP\Lambda}\bar \Lambda \gamma_5 N K^{-},\cr
\mathcal{ L}_{\gamma KK^*}&=&\frac{g_{\gamma KK^*}}{m_{K^*}}
\epsilon_{\mu \nu \alpha \beta} \partial^\nu A^\mu \partial^\beta
K^{*\alpha} K,\cr 
\mathcal{ L}_{\phi KK^*}&=&\frac{g_{\phi KK^*}}{m_{\phi}}
\epsilon_{\mu \nu \alpha \beta} \partial^\nu \phi^\mu \partial^\beta
K^{*\alpha}K,    
\end{eqnarray}
where the coupling constants and the cut-off masses are listed in
Table~\ref{tab:3}. 
\begin{table}[h]
  \begin{center}
    \caption{Coupling constants and cut-off masses used in box
      diagrams of Fig.~\ref{fig:4}}
    \label{tab:3}
    \begin{tabular}{lll}
\hline \hline
$g_{\gamma \rho \sigma}\hspace{1cm}$ {}\hspace{1cm}  {}
& 0.82{}\hspace{1cm}\hspace{1cm} {}
&Ref.\cite{Friman:1995qm}\\
$g_{\sigma NN}$&10.026& Ref.\cite{Friman:1995qm}\\
$g_{\pi NN}$&13.26& Ref.\cite{Friman:1995qm} \\
$g_{\pi \rho \phi}$ & -1.258 & Ref.\cite{PDG2012} \\
$g_{\phi \omega \sigma}$& -0.45 & Ref.\cite{PDG2012}\\
$g_{\gamma \omega \pi}$ & 0.557 & Ref.\cite{PDG2012} \\ 
$g_{\omega NN} $ &10.35& Ref.\cite{titov2002}\\
$g_{\rho NN}$&$ 3.72$& Ref.\cite{titov2002}\\
$g_{\phi KK}$&4.48&  Ref.\cite{PDG2012} \\
$g_{KN\Lambda}$&-13.26& Ref.\cite{nakayama2006} \\
$g_{\gamma KK^*}$&0.254 GeV${}^{-1}$&  Ref.\cite{PDG2012} \\
$g_{\phi KK^*}$&10.74&  Ref.\cite{PDG2012,Titov:2004xf}\\
\hline
$\kappa_{\omega}$&0&Ref.\cite{titov2002}\\
$\kappa_{\rho}$&6.1& Ref.\cite{titovKampfer2002}\\
\hline
$\Lambda_{\pi \rho \phi}$&1.05 GeV& Ref.\cite{Titov:2007fc} \\
$\Lambda_{\pi NN}$&1.05 GeV & Ref.\cite{Titov:2007fc}\\
$\Lambda_{\gamma \rho \sigma}$ &1.05 GeV &Ref.\cite{Friman:1995qm} \\
$\Lambda_{\sigma NN}$&1.1 GeV& Ref.\cite{Friman:1995qm} \\
$\Lambda_{\sigma}$&1 GeV& Ref.\cite{Friman:1995qm} \\
$\Lambda_{\sigma \rho \rho}$&0.9 GeV& Ref.\cite{Friman:1995qm} \\
$\Lambda_{\sigma\omega \phi}$&0.9 GeV& Ref.\cite{PDG2012}\\
$\Lambda_{\pi \gamma \omega }$&0.6 GeV& Ref.\cite{titov2002}\\
$\Lambda_{V} $ & 1.227 GeV &  Ref.\cite{sato1996} \\
$\Lambda_{K} $ & 1 GeV &   \\
\hline \hline
    \end{tabular}
  \end{center}
\end{table}
The invariant amplitudes for these box diagrams are derived as
follows: 
\begin{eqnarray}
  \label{eq:12}
\mathcal{ M}_{1 ,L}&=&
\frac{ g_{\gamma \rho \sigma}g_{\sigma NN}}{M_{\rho}
(t_{\sigma}-M_{\sigma}^2)} 
\left[ (k_1\cdot r)(\epsilon_{\gamma} \cdot \epsilon_\rho^*) 
- (k_1 \cdot \epsilon_\rho^*) (\epsilon_\gamma \cdot r) \right] 
\bar u(q) u(p_1) 
\bigg\{
\frac{ \Lambda_{\gamma \rho \sigma}^2-M_\sigma^2 } { t_\sigma - M_\sigma^2} 
\cdot 
\frac{\Lambda_{\sigma NN}^2-M_\sigma^2}{t_\sigma-M_\sigma^2} 
\bigg\},\cr
\mathcal{ M}_{1 ,R}&=&
\frac{-i g_{\phi \rho \pi} g_{\pi NN}}{M_{\phi}(t_{\pi}-M_{\pi}^2)} 
\epsilon_{\mu \nu\alpha \beta} \epsilon_{\phi}^{*\mu} k_2^\nu
\epsilon_\rho^\alpha r^\beta \bar u(p_2)\gamma_5 u(q) 
\times \bigg\{
\frac{ \Lambda_{\phi \rho \pi}^2-M_\pi^2 } { t_\pi - M_\pi^2} 
\cdot 
\frac{\Lambda_{\pi NN}^2-M_\sigma^2}{t_\pi-M_\pi^2} 
\bigg\},\cr
\mathcal{ M}_{2 ,L}&=& 
 \frac{-ig_{\gamma \omega \pi}g_{\pi NN}}
{M_{\omega}(t_\pi -M_\pi^2)}
\epsilon_{\mu \nu \alpha \beta}\epsilon_{\gamma}^{\mu} k_1^{\nu}
\epsilon_{\omega}^{*\alpha} r^{\beta}
\bar u(q)\gamma_5 u(p_1) \times
\left\{ \frac{\Lambda_{\gamma \omega \pi}^2 -M_{\pi}^2}
{t_{\pi}-M_{\pi}^2} \cdot
\frac{\Lambda_{\pi NN}^2 -M_{\pi}^2}
{t_{\pi}-M_\pi^2} \right\},\cr
\mathcal{ M}_{2,R}&=& 
 \frac{-i g_{\phi \omega \sigma} g_{\sigma NN}}
{M_{\phi}(t_\sigma -M_\sigma^2)}
\bar u(p_2) u(q) \big[ (r\cdot k_2)(\epsilon_\omega \cdot
\epsilon_{\phi}^*) 
-(r \cdot \epsilon_{\phi}^*)(k_2 \cdot \epsilon_\omega) \big]
\cr
&&\times
 \left\{
\frac{\Lambda_{\phi \omega \sigma}^2 -M_{\sigma}^2}
{t_{\sigma}-M_\sigma^2}\cdot
\frac{\Lambda_{\sigma NN}^2 -M_{\sigma}^2}
{t_{\sigma}-M_\sigma^2}   \right\},\cr
\mathcal{ M}_{3,L}&=& 
\frac{g_{\rho NN}g_{\gamma \rho \sigma}}{M_{\rho}(t_\rho -M_\rho^2)}
\Big[k_1^\alpha (\epsilon_\gamma \cdot r)
-\epsilon_{\gamma}^{\alpha} (k_1 \cdot r) \Big]
\bar u(p_2)\bigg[  \gamma_\alpha (1+\kappa_\rho)
-\frac{\kappa_\rho}{M_p}q^\alpha \bigg]u(p_1)\cr
&& \times \Bigg\{ \Bigg(
\frac{\Lambda_{ \rho}^2}
{\Lambda_{ \rho}^2 -(\bm k_1 -\bm r)^2}
\Bigg)^2
\Bigg\},\cr
\mathcal{ M}_{3, R}&=& 
\frac{g_{\omega NN}g_{\phi \omega \sigma }}{M_{\phi}(t_\omega -M_\omega^2)}
\Big[ (r \cdot \epsilon_{\phi}^{*})k_2^{\mu} 
- (r \cdot k_2 -M_{\phi}^2) \epsilon_{\phi}^{*\mu}
\Big]\cr
&&\times
\bar u(p_1) \bigg[ \gamma_\mu (1+\kappa_\omega)
-\frac{\kappa_\omega}{2M_p} 
q_\mu \bigg] u(q)  
\Bigg\{
\Bigg(
\frac{\Lambda_{ \omega}^2}
{\Lambda_{ \omega}^2 -(\bm r -\bm k_2)^2}
\Bigg)^2
\Bigg\},\cr
\mathcal{ M}_{4,L}&=& 
\frac{-g_{\omega NN}g_{\gamma \omega \pi}}
{M_{\omega} (t_\omega -M_{\omega}^2)}
\epsilon_{\mu \nu \alpha \beta} \epsilon_{\gamma}^{\mu} k_1^{\nu}r^{\beta} 
\bar u(q) \left[  \gamma^\alpha (1+\kappa_\omega)
-\frac{\kappa_\omega}{M_P} q^{\alpha}
\right]u(p_1) \cr
&& \times 
\Bigg\{
\Bigg(
\frac{\Lambda_{ \omega}^2}
{\Lambda_{ \omega}^2 -(\bm r -\bm k_2)^2}
\Bigg)^2
\Bigg\}, \cr
\mathcal{ M}_{4 ,R}&=& 
\frac{-g_{\rho NN}g_{\phi \rho \pi}}{M_\phi (t_\rho -M_\rho^2)}
\epsilon_{\mu \nu \alpha \beta} \epsilon_{\phi}^{*\mu} k_2^{\nu}
r^\beta \bar u (p_2)  
\left[ \gamma^\alpha (1+\kappa_\rho) - 
\frac{\kappa_\rho}{M_N} ql^\alpha \right]u(q), \cr
&&\times 
\Bigg\{
\Bigg(
\frac{\Lambda_{ \rho}^2}
{\Lambda_{ \rho}^2 -(\bm k_1 -\bm r)^2}
\Bigg) ^2
\Bigg\} ,\cr
\mathcal{ M}_{5 ,L}&=&
 \frac{-2ieg_{KP\Lambda}}{(t_L -M_K^2)}
(r \cdot \epsilon_\gamma)  \bar u(q) \gamma_5 u(p_1)  
\times \bigg\{
\left( \frac{\Lambda_{K}^2 -M_{K}^2}
{t_{K}-M_K^2} \right)^2  \bigg\},\cr
\mathcal{ M}_{5,R}&=& 
 \frac{2ig_{\phi KK}g_{KP\Lambda}}{(t_R^2-M_K^2)}
(r\cdot \epsilon_\phi^*) \bar u(p_2)\gamma_5 u(q) 
\times \bigg\{
\left( \frac{\Lambda_{K}^2 -M_{K}^2}
{t_{K}-M_K^2} \right)^2  \bigg\},\cr
\mathcal{ M}_{6 ,L}&=& 
\frac{-ig_{\gamma K K^*}g_{KP\Lambda}}
{M_{K^*} (t_L-M_K^2)}
\epsilon_{\mu \nu \alpha \beta}  \epsilon_{\gamma \mu}  k_{1}^{\nu}
 \epsilon_{K^*}^{\alpha} r^\beta  \bar u(q) \gamma_5 u(p_1) 
\times \bigg\{
\left( \frac{\Lambda_{K}^2 -M_{K}^2}
{\Lambda_{K}^2-t_{K}} \right)^2  \bigg\},\cr
\mathcal{ M}_{6,R}&=& 
 \frac{i g_{\phi  K K^*}g_{KP\Lambda}}
{M_{\phi}(t_R-M_K^2)}
\epsilon_{\mu \nu \alpha \beta}\epsilon_{\phi}^{*\mu} k_2^{\nu}
\epsilon_{K^*}^{\alpha}  r^\beta \bar u(p_2) \gamma_5 u(q)
\times \bigg\{
\left( \frac{\Lambda_{K}^2 -M_{K}^2}
{\Lambda_{K}^2-t_{K}} \right)^2  \bigg\},   
\end{eqnarray}
where the subscripts $1,\cdots 6$ correspond to the box diagrams
appearing in Fig.~\ref{fig:4} in order. The other subscripts $L$ and
$R$ denote the $\gamma p\to MB$ and $MB\to \phi p$ parts,
respectively. In Figs.~\ref{fig:5} and \ref{fig:6} we draw the results
of the total cross sections for the $\gamma p\to \rho p$ and $\gamma p
\to \omega p$ reactions, respectively. The results are qualitatively
in agreement with the experimental data.

\section{Numerical results and discussion\label{sec:3}}
\begin{figure}[ht]
\centerline{
       \includegraphics[width=13cm]{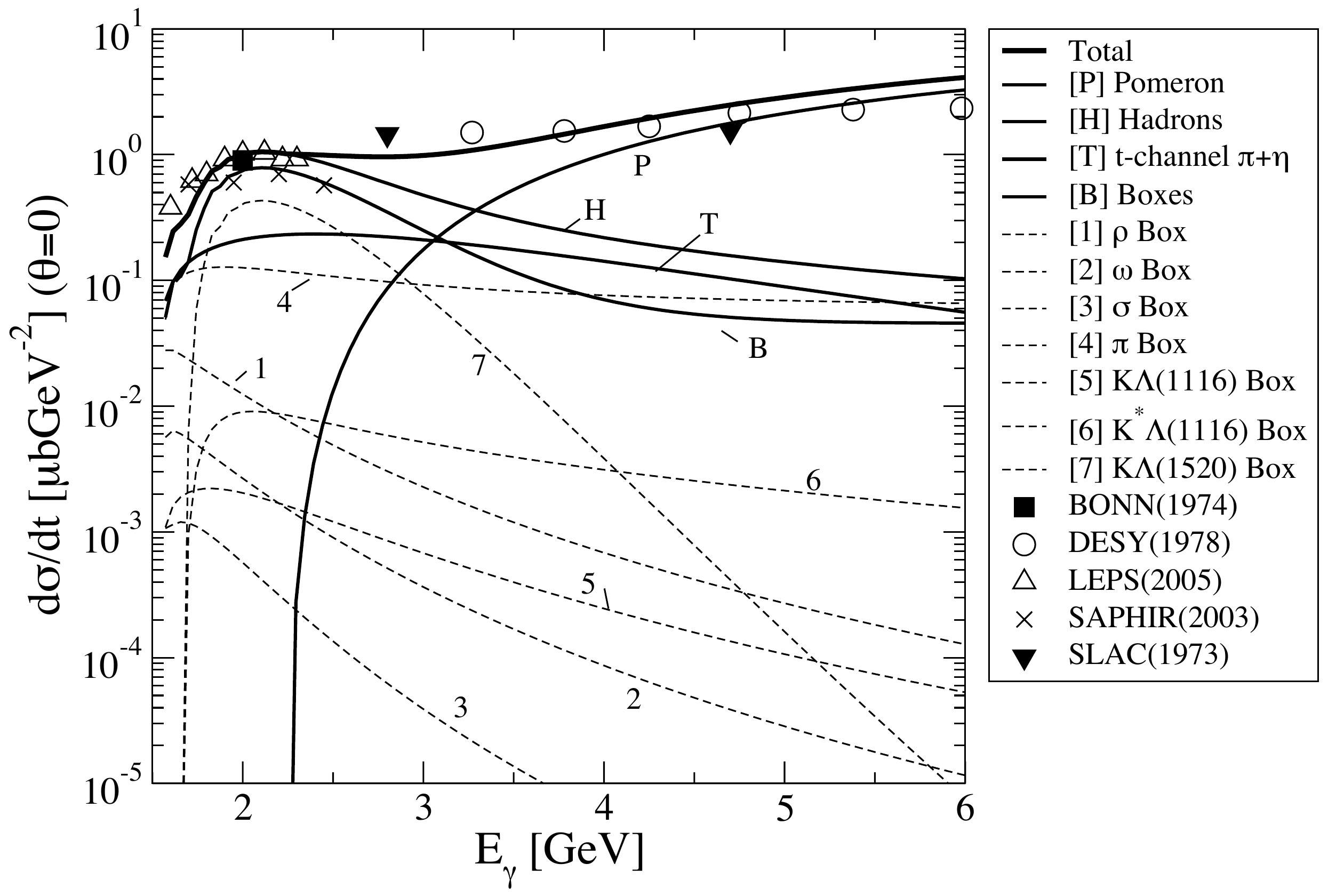}
}
\caption{Differential cross section as a function of the photon
  energy $E_{\gamma}$. The thick solid curve depicts the result with
  all contributions included. The solid curves with the symbols $P$,
  $T$, $B$ and $H$ represent the Pomeron contribution, those of $\pi$-
  and $\eta$-exchanges, those of all the box diagrams, and the total
  contribution of hadronic diagrams ($T+B$), respectively. The dashed
  curves with numbers in order denote the effects of the seven box
  diagrams separately. }    
\label{fig:7}  
\end{figure}
We are now in a position to discuss the numerical results for $\phi$
photoproduction. We start with  the differential cross section at the
forward angle $d\sigma/dt(\theta=0)$ as a function of the photon
energy $E_{\gamma}$ in the laboratory frame. The parameters are
determined in the following manner. Since the the Pomeron-exchange in
the low-energy region is not much understood, we fit the parameter for
the overall strength $C_p$ and that for the threshold $s_{\mathrm{th}}$
in Eq.(\ref{eq:4}) in such a way that the Pomeron-exchange reproduces
the high energy behavior of the differential cross section:
$C_p=8\,\mathrm{GeV}^{-1}$ and $s_{\mathrm{th}}=3.83\,\mathrm{GeV}^2$.  
On the other hand, We fix the cut-off parameters for the
$K\Lambda^*(1520)$ box diagrams to describe the $E_\gamma$ dependence
of $d\sigma/dt$ in lower energy region, in particular, to explain the
well-known bum-like structure around $E_\gamma\approx 2.3$ GeV. The
parameters of all other hadronic diagrams are taken from existing
references as explained in the previous section.  

Figure~\ref{fig:7} illustrates various contributions to
$d\sigma/dt(\theta=0)$ as a function of the 
photon energy $E_\gamma$ from the Pomeron-exchange, the $t$-channel
$\pi$- and $\eta$ exchanges, and seven box diagrams. The solid curve with
symbol $P$ draws the contribution of the the Pomeron-exchange to
$d\sigma/dt$. As expected, it governs $E_\gamma$ dependence in the
higher energy region ($E_{\gamma} \geq 3$GeV). Note, however, that the
Pomeron does not contribute to $d\sigma/dt$ below around
$E_\gamma=2.3\, \mathrm{GeV}$ in the present work. The $\pi$- and 
$\eta$-exchanges provide a certain amount of effects on the
differential cross section (solid curve with symbol $T$). 
The contribution of the $\pi$- and $\eta$-exchanges start to increase
from the threshold energy and then it decreases very slowly when it
reaches approximately 3 GeV. Thus, the effects of the $\pi$- and
$\eta$-exchanges are quite important in the lower $E_\gamma$ energy
region up to $3$ GeV, where the Pomeron-exchange overtakes the 
$\pi$- and $\eta$-exchanges. 

Except for the $K\Lambda^*(1520)$ box
diagram, all other box contributions turn out to be negligibly
small. However, the $K\Lambda^*(1520)$ box diagram plays an essential
role in describing the experimental data for $d\sigma/dt$ in the lower
$E_\gamma$ region, in particular, in explaining the bump-like
structure near $2.3$ GeV. This is very different from the conclusion
of Ref.~\cite{ozaki2009}, where the $K\Lambda^*(1520)$ seems to be
suppressed in the $K$-matrix formalism. The reason lies in the fact
that we have introduced different form factors for the $\gamma p\to
K\Lambda^*$ and $K \Lambda^*\to \phi p $ reactions. In general, form
factors are given as functions of two Mandelstam variables for the box
diagrams, i.e. $F(s,\,t)$, since we have two off-shell particles in
the $s$-channel and other two off-shell particles in the $t$-channel.     
However, it is very difficult to preserve the gauge invariance in the
presence of the form factors. Thus, we have introduced a type of
overall form factors to keep the gauge invariance in the $\gamma p \to
K\Lambda^*$ part, as written in Eq.(\ref{eq:11}). To keep the
consistency, we also have included a similar type of the form factors
in the $K\Lambda^*\to \phi p$ part. With these form factors
considered, we find that the $K\Lambda^*$ box diagram is indeed
enhanced as shown in Fig.~\ref{fig:7} in comparision with
Ref.~\cite{ozaki2009}. The contribution of the $K\Lambda^*$ box
diagram increases sharply up to $E_\gamma \approx 2$ GeV and
then falls off linearly. The result of the  $K\Lambda^*$ box diagram
indicates that the off-shell effects, which arise from the form
factors and the rescattering equation, may come into play.

Considering the fact that the $K^*\Lambda$ threshold energy
($E_{\mathrm{th}}\approx 
2$ GeV) is very close to that of $\phi$ photoproduction 
($E_{\mathrm{th}}\approx1.96$ GeV), one might ask why the contribution
of the $K^*\Lambda$ is suppressed. While the $K\Lambda^*(1520)$
channel ($E_{\mathrm{th}}\approx 2$ GeV) is directly related to $\phi
p$, since both are the subreactions of the $\gamma p\to K\bar{K}p$
process, the $\gamma p\to K^*\Lambda$ reaction is distinguished from
those two reactions, because the $K^*\Lambda$ channel is related to 
$\gamma p \to \pi K \Lambda$ reaction. Thus, one can qualitatively
understand why the contribution of the $K^*\Lambda$ box diagrams is
suppressed.  

\begin{figure}[ht]
 \centerline{
       \includegraphics[width=13cm]{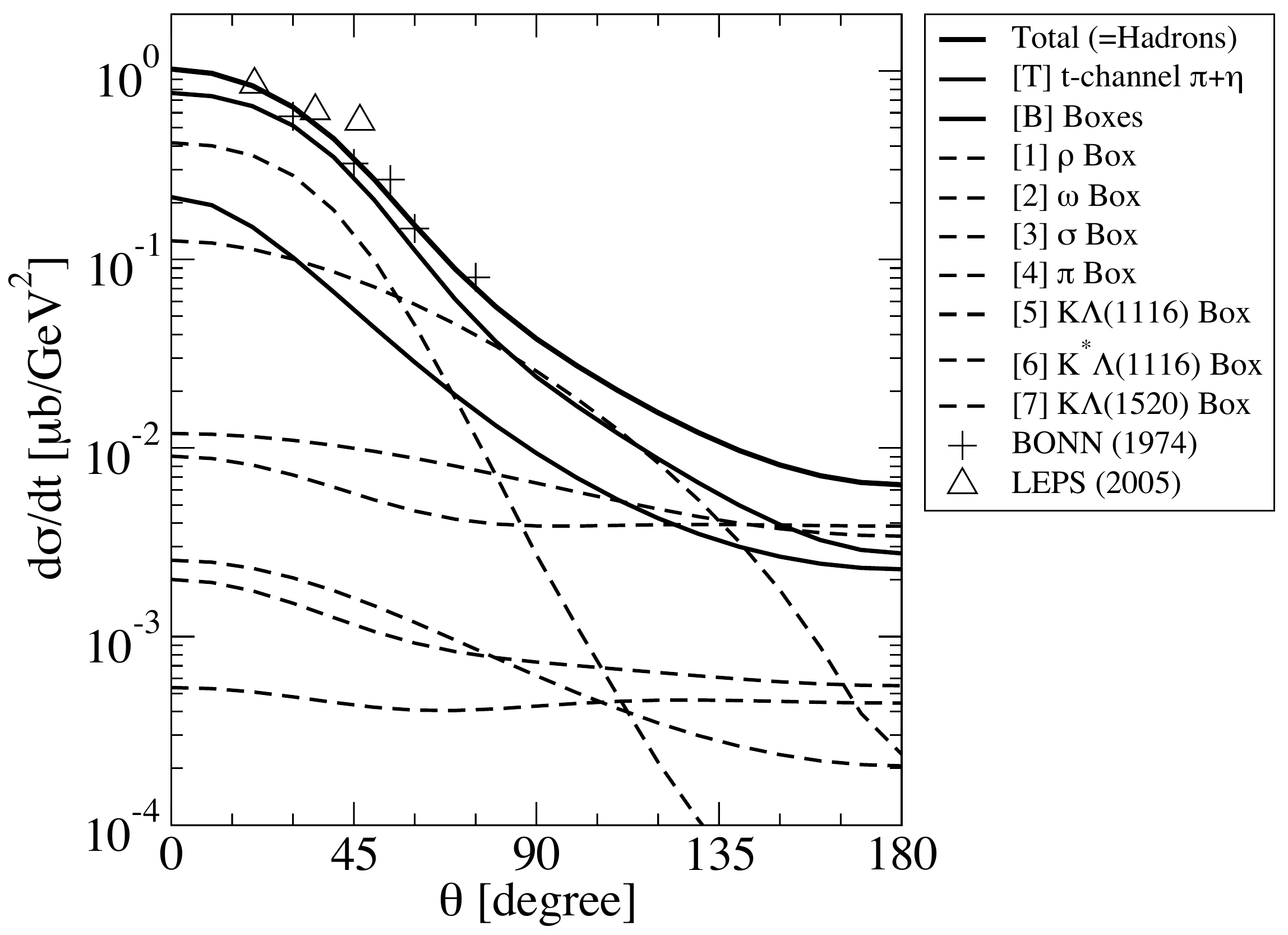}
}
  \caption{The differential cross section as a function of the
    scattering angle $\theta$ with the photon energy at 
    $E_{\mathrm{\gamma}}=2$ GeV. The thick solid curve depicts the result with
  all hadronic contributions included. The solid curves with the
  symbols $T$ and $B$ represent the contribution of the $\pi$-
  and $\eta$-exchanges and those of all the box diagrams,
  respectively. The dashed curves with numbers in order denote the
  effects of the seven box diagrams separately.} 
  \label{fig:8}
\end{figure}
In Fig.~\ref{fig:8}, the differential cross section as a function of the
scattering angle is depicted at $E_\gamma = 2$ GeV. Since the
Pomeron-exchange is suppressed at this  
photon energy because of $s_{\mathrm{th}}=2.3$ GeV, we can examine
each hadronic contribution to the differential cross section more in
detail. Figure~~\ref{fig:8} clearly shows that the $K\Lambda(1520)$
box diagram is the most dominant one among the hadronic
contributions. Adding all the effects of the box diagrams, we find
that the box contributions almost describe the $\theta$
dependence. Together with the $\pi$- and $\eta$-exchanges, the result
of the differential cross section is in good agreement with the
experimental data~\cite{Mibe:2005er,BESCH:1974aa}. 

\begin{figure}[ht]
\centerline{
       \includegraphics[width=10cm]{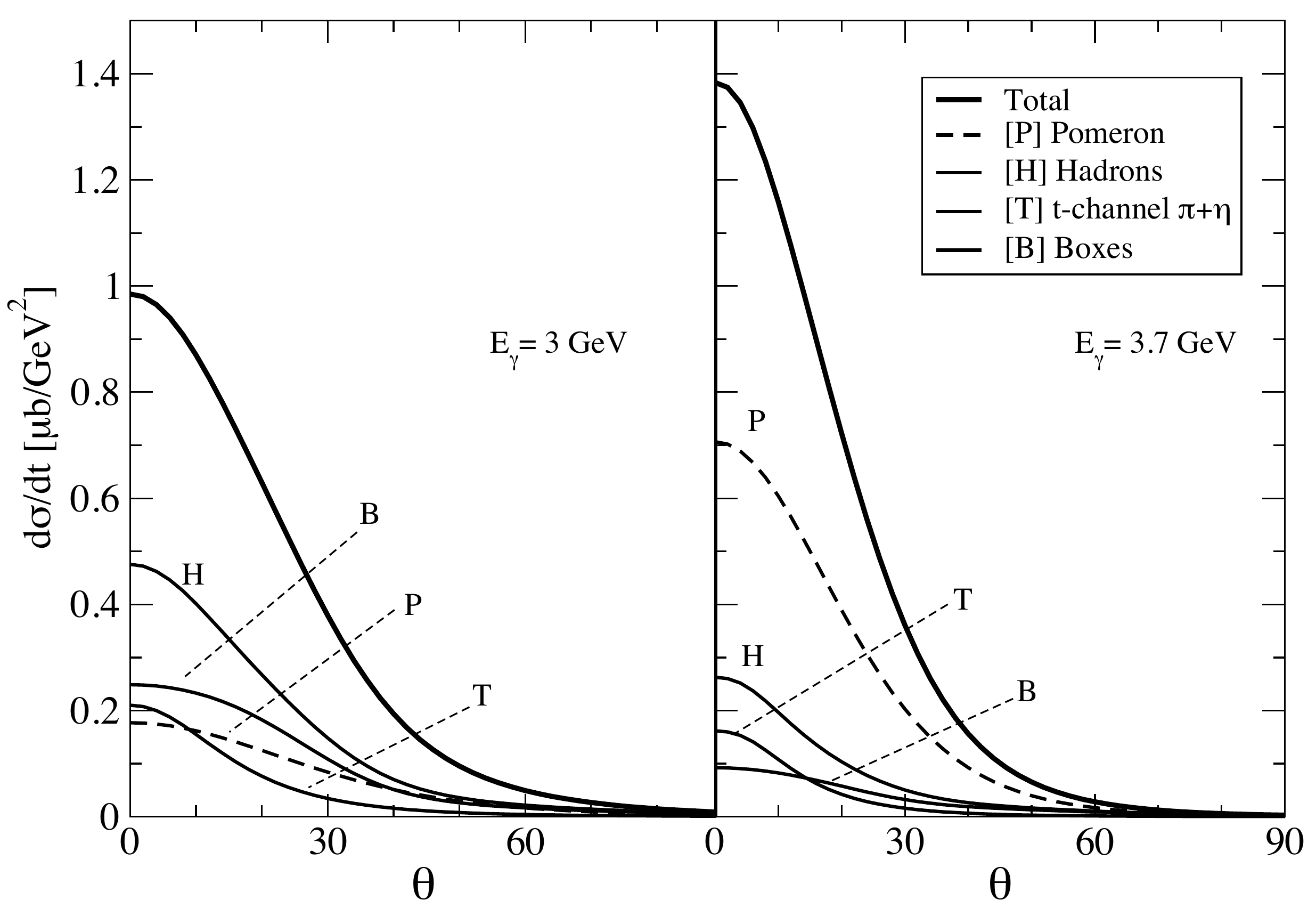}
}
  \caption{The differential cross section as a function of the
    scattering angle $\theta$ with two different photon energies
    $E_{\mathrm{\gamma}}=3$ GeV and $3.7$ GeV. The thick solid curve
    depicts the result with all contributions included. The solid
    curves with the symbols $P$, $T$, $B$ and $H$ represent the
    Pomeron contribution, those of $\pi$- and $\eta$-exchanges, those
    of all the box diagrams, and the total contribution of hadronic
    diagrams ($T+B$), respectively. 
    } 
  \label{fig:9}
\end{figure}
The differential cross section as a function the scattering angle are
drawn in Fig.~\ref{fig:9}. The left and right panels correspond to the
photon energies $E_{\mathrm{\gamma}}=3$ and $3.7$~GeV, respectively. As
expected, the hadronic contribution is dominant over the
Pomeron-exchange at the lower photon 
energy, while at $E_\gamma=3.7$ GeV, the Pomeron governs the $\gamma
p\to \phi p$ process. Interestingly, the effects of the box diagrams,
in particular, the $K\Lambda^*(1520)$ one, 
turn out to be larger than those of the $\pi$- and $\eta$-exchanges,
whereas the box diagrams seem to be suppressed at higher photon
energies. It implies that the $K\Lambda^*(1520)$ box diagram
influences $\phi$ photoproduction only in the vicinity of the
threshold energy. 
\begin{figure}[h]
  \begin{center}       
       \includegraphics[width=14cm]{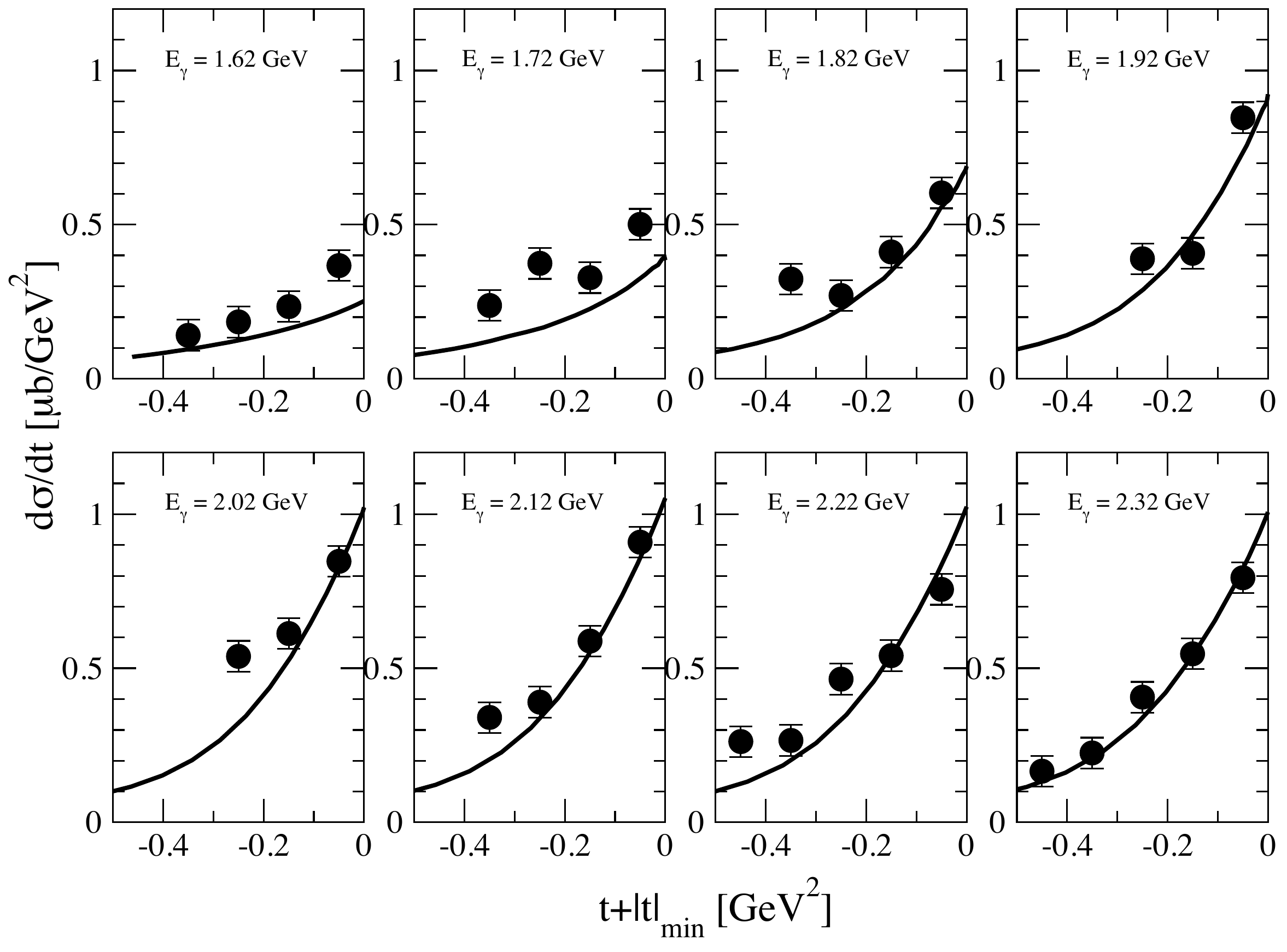}
  \end{center}
  \vspace{-10pt}
  \caption{Differential cross sections of the $\gamma p \to \phi p$
    reaction as a function of $t+|t|_{\mathrm{min}}$ with eight
    different photon energies. The experimental data are taken from
    Ref. \cite{Mibe:2005er}.}   
  \label{fig:10}
\end{figure} 
Figure~\ref{fig:10} depicts the results of the
differential cross section as a function of $t+|t|_{\mathrm{min}}$
with eight different photon energies,
where $|t|_{\mathrm{min}}$ is the minimum 4-momentum transfer from the
incident photon to the $\phi$ meson. The results are in good agreement 
with the experimental data taken from the measurement of the LEPS
collaboration~\cite{Mibe:2005er}.  

It is of great importance to examine the angular distribution of
the $\phi\to K^+ K^-$ decay in the $\phi$ rest frame or in the 
Gottfried-Jackson (GJ) frame, since it makes the helicity amplitudes 
accessible to experimental
investigation~\cite{Gottfried:1964nx,Schilling:1969um}. The 
detailed formalism for the angular distribution of the $\phi$ meson
decay can be found in Refs.~\cite{Schilling:1969um,Titov:1999eu}.  
The decay angular distribution of $\phi$ photoproduction was measured
at SAPHIR/ELSA~\cite{Barth:2003bq} but the range of the photon energy
is too wide. On the other hand, the LEPS collaboration measured the
decay angular distribution at forward angles ($-0.2
<t+|t|_{\mathrm{min}}$) in two different energy regions:
$1.97<E_\gamma< 2.17$ GeV and $2.17 <E_\gamma<2.37$ 
GeV~\cite{Mibe:2005er}, which are related to the energy around the
local maximum of the cross section and that above the local maximum,
respectively. Therefore, we have computed the decay angular
distributions at two photon energies, i.e. $E_\gamma=2.07$ GeV and
$E_\gamma = 2.27$ GeV, which correspond to the center values of the
given ranges of $E_\gamma$ in the LEPS experiment.
   
\begin{figure}[h]
  \begin{center}       
       \includegraphics[width=12.5cm]{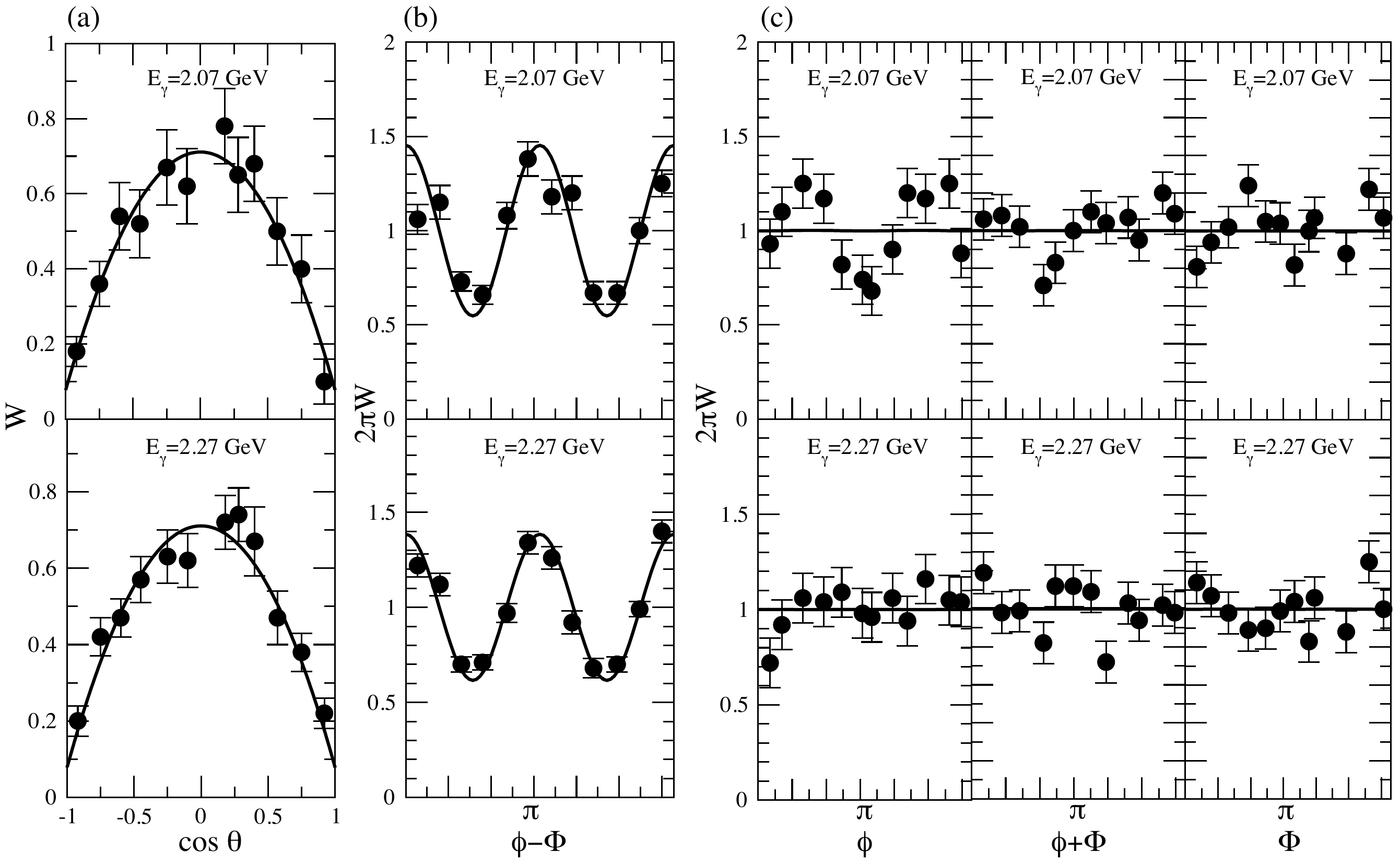}
\caption{
The decay angular distributions for
$-0.2<t+|t|_{\mathrm{min}}$ in the Gottfried-Jackson frame. We take 
the center values of the energy ranges measured by the LEPS
collaboration~\cite{Mibe:2005er}, i.e. $E_\gamma=2.07$ GeV and
$E_{\gamma}=2.27$ GeV.  The experimental data are taken from
Ref.~\cite{Mibe:2005er}.  
} 
 \label{fig:11}      
  \end{center}
\end{figure}

The one-dimensional decay angular
distributions $W(\cos\theta)$, $W(\phi-\Phi)$, $W(\phi)$
are presented in Fig.~\ref{fig:11}, which are expressed respectivley
as      
\begin{eqnarray}
  \label{eq:16-1}
W(\cos \theta)&=& \frac{1}{2}(1-\rho^{0}_{00})+
\frac12 \left(3\rho^{0}_{00}- 1 \right)\cos^2 \theta ,\cr 
\label{eq:16-2}
 2\pi W(\phi -\Phi)&=&1+2p_\gamma  \overline{\rho}^{1}_{1-1}
\cos 2(\phi -\Phi),\cr
 \label{eq:16-3} 
2\pi W(\phi)&=&1-2\mathrm{Re} \rho^{0}_{1-1} \cos 2 \phi ,\cr
 \label{eq:16-4}
2\pi W(\phi +\Phi)&=&1+2p_\gamma  \Delta_{1-1}
\cos 2(\phi +\Phi),\cr
\label{eq:16-5}   
2\pi W(\Phi)&=& 1+2p_\gamma  \rho' \cos 2\Phi,   
\end{eqnarray}
where $\theta$ and $\phi$ denote the polar and azimuthal
angles of the decay particle $K^+$ in the GJ frame. $\Phi$ represents 
the azimuthal angle of the photon polarization in the center-of-mass  
frame. $P_\gamma$ stands for the degree of polarization of the photon
beam. $\overline{\rho}^{1}_{1-1}$, $\Delta_{1-1}$, and $\rho'$ are
defined as  
\begin{eqnarray}
  \label{eq:17-1} 
\overline{\rho}^{1}_{1-1} &=& 
\frac{1}{2}(\rho^{1}_{1-1}-\mathrm{Im} \rho^{2}_{1-1}),\cr
\Delta_{1-1} &=&
\frac{1}{2}(\rho^{1}_{1-1}+\mathrm{Im} \rho^{2}_{1-1}),\cr
\rho' &=& 2\rho^{1}_{11}+\rho^{1}_{00} . 
\end{eqnarray} 
The expressions for the spin-density matrix elements
$\rho_{\lambda\lambda'}^\alpha$ with the helicities $\lambda$ and
$\lambda'$ for the $\phi$ meson can be found in Appendix~\ref{app:a} . 

The panel (a) of Fig.~\ref{fig:11} draws the one-dimensional decay
polar-angle distributions $W(\cos\theta)$.
As pointed out by Refs.~\cite{Mibe:2005er,Chang:2010dg}, the decay
distribution behaves approximately as $\sim(3/4) \sin^2\Psi$, which
indicates that the helicity-conserving processes are dominant as shown
in Eq.(\ref{eq:16-1}). It means that $t$-exchange particles with
unnatural parity at the tree level do not contribute to
$W(\cos\theta)$. As will be discussed later, $\rho_{00}^0$ from the
$\pi$- and $\eta$-exchanges, which is related to the single spin-flip 
amplitude in the GJ frame, exactly vanishes. On the other hand, all
hadronic box diagrams contribute to it. Though the Pomeron-exchange
might contribute to this spin-density matrix element, it does not play
any role below 2.3 GeV.
The panel (b) of Fig.~\ref{fig:11} shows the results of
$W(\phi-\Phi)$, which are in good agreement with the LEPS data,
whereas the panel (c) depicts those of $W(\phi)$, $W(\phi+\Phi)$, and
$W(\Phi)$, respectively, which deviate from the data. In fact, the
data show somewhat irregular behavior which does not seem to be easily
reproduced.  

\begin{table}[ht]
\caption{$\phi$ density matrix in the forward scattering at
  $E_{\mathrm{\gamma}}=2$ GeV} 
\label{tab:4}

\begin{center}
\begin{tabular}{cccccc}
\hline
\hline
&$\rho^{0}_{00}$
&$\overline{\rho}^{1}_{1-1}$ 
&$\mathrm{Re}\rho^{0}_{1-1}$
&$\Delta_{1-1}$
&$\rho'$ \tabularnewline    
\hline
$t$-channel $\pi^0+\eta$ &0&-0.5&0&0  &0 \tabularnewline    
\hline
$\rho$ box 
& {\footnotesize $0.651$} 
&{\footnotesize $-0.175$} 
&{\footnotesize $2.97\times 10^{-4}$}
&{\footnotesize $-8.94\times 10^{-6}$}
&{\footnotesize $1.37\times 10^{-2}$} \tabularnewline    
$\omega$ box 
& {\footnotesize $0.035$} 
&{\footnotesize $-0.48$} 
&{\footnotesize $9.26\times 10^{-4}$}
&{\footnotesize $-8.72\times 10^{-7}$}
&{\footnotesize $-1.05\times 10^{-3}$} \tabularnewline    
$\sigma$ box 
& {\footnotesize $0.254$} 
&{\footnotesize $-0.066$} 
&{\footnotesize $-8.85\times 10^{-3}$}
&{\footnotesize $2.03\times 10^{-4}$}
&{\footnotesize $-7.93\times 10^{-4}$} \tabularnewline    
$\pi$ box 
& {\footnotesize $0.061$} 
&{\footnotesize $0.448$} 
&{\footnotesize $5.57\times 10^{-4}$}
&{\footnotesize $1.79\times 10^{-4}$}
&{\footnotesize $1.15\times 10^{-3}$} \tabularnewline    
$K \Lambda(1116)$ box 
& {\footnotesize $0.025$} 
&{\footnotesize $0.488$} 
&{\footnotesize $-1.08\times 10^{-2}$}
&{\footnotesize $7.85\times 10^{-5}$}
&{\footnotesize $-2.21\times 10^{-2}$} \tabularnewline    
$K^* \Lambda(1116)$ box 
& {\footnotesize $0.030$} 
&{\footnotesize $0.485$} 
&{\footnotesize $1.39\times 10^{-3}$}
&{\footnotesize $1.10\times 10^{-6}$}
&{\footnotesize $2.06\times 10^{-3}$} \tabularnewline    
$K^+ \Lambda(1520)$ box 
&{\footnotesize $3.1 \times 10^{-4}$}
&{\footnotesize $0.499$}
&{\footnotesize $-2.95\times 10^{-3}$}
&5.131$\times 10^{-6}$ 
&-6.02$\times 10^{-3}$ \tabularnewline    
\hline
box all 
& {\footnotesize $6.62 \times 10^{-2}$} 
&{\footnotesize $0.455$} 
&{\footnotesize $2.46\times 10^{-4}$}
&{\footnotesize $1.74\times 10^{-4}$}
&{\footnotesize $5.69\times 10^{-4}$} \tabularnewline    
hadrons 
& {\footnotesize $5.13 \times 10^{-2}$} 
&{\footnotesize $0.24$} 
&{\footnotesize $5.64\times 10^{-4}$}
&{\footnotesize $1.34\times 10^{-4}$}
&{\footnotesize $-1.99\times 10^{-4}$} \tabularnewline    
\hline\hline
\end{tabular}
\end{center}
 \end{table}
As shown in Fig.~\ref{fig:11}, the decay angular distributions shed
light on the production mechanism of the $\phi$ meson, since they
make it possible to get access experimentally to the spin-density
matrix elements, or the helicity amplitudes of $\phi$
photoproduction. It has important physical implications, because even 
though some diagrams seem to contribute negligibly to the cross
sections, they might have definite effects on the decay angular
distributions. In Table~\ref{tab:4}, The contributions of each box
diagram to the various spin-density matrix elements at $E_\gamma =
2\,\mathrm{GeV}$ are listed. As expected, the $\pi$- and
$\eta$-exchanges contribute only to $\overline{\rho}_{1-1}^1$. The
hadronic box diagrams mainly 
contribute to $\rho_{00}^0$ and $\overline{\rho}_{1-1}^1$ and are
almost negligible to other components. Interestingly, the $\rho p$ box 
diagram is the dominant one for $\rho_{00}^0$, even though it provides
much smaller effects on the differential cross section than the
$K\Lambda^*(1520)$ one.

\begin{figure}[h]
  \begin{center}       
       \includegraphics[width=12.5cm]{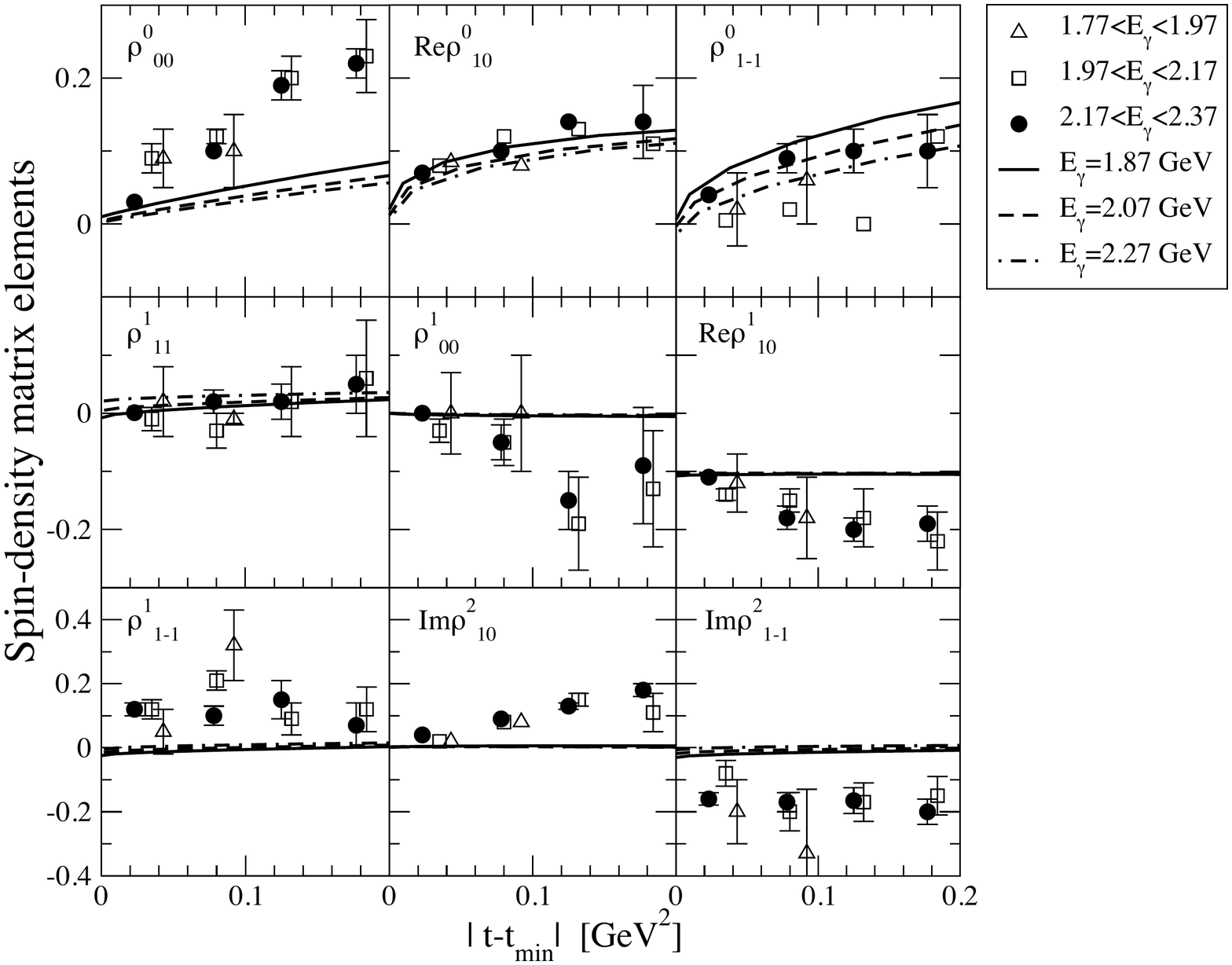}
\caption{The density matrix elements as a function of
$|t-t_{\mathrm{min}} |$ for three different photon energies,
i.e. $1.87$ GeV, $2.07$ GeV, and $2.27$ GeV, to which   
the solid, dotted, and dot-solid curves correspond.
The experimental data with three different ranges of the photon energy
are taken from Ref.~\cite{Chang:2010dg}.    
} 
 \label{fig:12}      
  \end{center}
\end{figure}  
Rcently, the LEPS experiment measured the spin-density matrix 
elements for $\gamma p \to \phi p$~\cite{Chang:2010dg} in the range of 
$E_\gamma=1.6-2.4$ GeV in which the Pomeron-exchange does not play any  
important role, in particular, in the present approach. Thus, we can
examine the hadronic contributions to each spin-density matrix
elements. Figure~\ref{fig:12} illustrates the various spin-density
matrix elements, compared with the LEPS data. Since the experimental
data are given in the finite range of $E_\gamma$, we just take the
three center values corresponding to the ranges,
i.e. $E_\gamma=1.87,\,2.07,\,2.27$ GeV. The hadronic diagrams
considered in the present work describe quantitatively
$\mathrm{Re}\rho_{10}^0$, $\rho_{1-1}^0$ and $\rho_{11}^1$. However,
the deviations are found in other spin-density matrix elements as
$t-|t|_{\mathrm{min}}$ increases. 

\section{Summary and outlook}
In the present work, we aimed at investigating the coupled-channel
effects arising from the hadronic intermediate box diagrams to $\phi$
photoproduction near the threshold region in addition to the Pomeron-,
$\pi$-, and $\eta$-exchanges. In particular, the bump-like structure
near $E_\gamma\approx 2.3$ GeV, which was reported by the LEPS
experiment~\cite{Mibe:2005er}, sheds light on the production mechanism
of the $\phi$ meson in the vicinity of the threshold, since the
Pomeron-exchange was shown to be not enough to explain this peculiar
structure of $\phi$ photoproduction. Thus, we studied in detail the
effects of the seven different box diagrams such as $\rho N$, $\omega
N$, $\sigma N$, $\pi N$, $K\Lambda(1116)$, $K^*\Lambda(1116)$, and
$K\Lambda(1520)$. In order to take into account the rescattering
terms, we employed the Landau-Cutkosky rule in dealing with these box
diagrams.

Since it turned out that the $K\Lambda^*(1520)$ box diagram played a
dominant role among hadronic contributions in the lower-energy region,
we scrutinized its contribution to $\phi$ photoproduction. We
introduced the form factors depending on both the $s$ and $t$
Mandelstam variables in such a way that the total cross section of the
$\gamma p\to K \Lambda^*(1520)$ reaction was well reproduced. All
other box diagrams were constructed by utilizing the previous
theoretical works and by reproducing the corresponding experimental
data when they were available. We examined each contribution carefully
by computing the differential cross section of $\phi$
photoproduction. While the $K\Lambda^*$ box diagram 
was found to be the most dominant near the 2 GeV, all other box
diagrams turned out to be very small. The results were in good
agreement with the LEPS data including the bump-like structure. We
also computed the differential cross section as a function of
$t+|t|_{\mathrm{min}}$ and found it to be in good agreement with the
experimental data.  

We investigated the contributions of hadronic box diagrams to the
decay angular distributions. While the one-dimensional angular
distributions $W(\cos\theta)$ and $W(\phi-\Phi)$ were 
in good agreement with the experimental data, other three angular
distributions seemed to deviate from the LEPS experimental data. We
also examined the various spin-density matrix elements, which were
measured recently by the LEPS collaboration. We found that the
hadronic box diagrams describe the experimental data for
$\mathrm{Re}\rho_{10}^0$, $\rho_{1-1}^0$ and $\rho_{11}^1$ were well
reproduced. While the present results explain near
$t-|t|_{\mathrm{min}}\approx 0$ relatively well for other spin-density
matrix elements, they deviated from the expeimental data as
$t-|t|_{\mathrm{min}}\approx 0$ increased.

As shown in the present work, the intermediate box diagrams, in
particular, the $K\Lambda^*(1520)$ one, play crucial roles in
explaining the cross sections of the $\gamma p\to \phi p$ reaction in
the vicinity of the threshold. Other box diagrams also provided certain
effects on the part of the spin-density matrix elements. We have
considered in this work only the imaginary part of the transition
amplitudes of the box diagrams based on the Landau-Cutkosky
rule. However, the results of the spin-density matrix elements already
indicate that we should carry out a theoretical analysis of $\phi$
photoproduction more systematically and quantitatively. Thus, we need
to investigate a full coupled-channel formalism and to solve
rescattering equations with the real parts of the box diagrams fully
taken into account. Another interesting and important problem
is to extend our approach to the neutron target, 
since some of considered amplitudes are isospin-dependent. The 
corresponding works are under way.   

\section*{Acknowledgments}
The authors are grateful to S.~Ozaki for valuable discussions. 
A.H. is supported in part by the Grant-in-Aid for Scientific Research
on Priority Areas titled ``Elucidation of New Hadrons with a Variety
of Flavors''(E01:21105006).  The work of H.Ch.K. was supported by
Basic Science  Research Program through the National Research
Foundation of Korea funded by the Ministry of Education, Science and
Technology (Grant Number: 2012001083). 
\appendix
\section{the spin-density matrix elements~\label{app:a}}
The spin-density matrix elements are expressed in terms of the helicity
amplitudes~\cite{Schilling:1969um,Titov:1999eu} 
\begin{eqnarray}
  \label{eq:app1}
 \rho_{\lambda \lambda'}^{0} &=& 
 \frac{1}{N} \sum_{ \lambda_{\gamma}, \lambda_{i},\lambda_{f}}
 T_{\lambda_f , \lambda ; \lambda_i ,\lambda_{\gamma}}
 T_{\lambda_f , \lambda' ; \lambda_i , \lambda_{\gamma}}^{*} ,  \cr
 \rho_{\lambda \lambda'}^{1} &=& 
 \frac{1}{N} \sum_{ \lambda_{\gamma}, \lambda_{i},\lambda_{f}}
 T_{\lambda_f , \lambda ; \lambda_i , -\lambda_{\gamma}}
 T_{\lambda_f , \lambda' ; \lambda_i , \lambda_{\gamma}}^{*} ,  \cr
 \rho_{\lambda \lambda'}^{2} &=& 
 \frac{i}{N} \sum_{ \lambda_{\gamma}, \lambda_{i},\lambda_{f}}
\lambda_\gamma T_{\lambda_f , \lambda ; \lambda_i , -\lambda_{\gamma}}
 T_{\lambda_f , \lambda' ; \lambda_i , \lambda_{\gamma}}^{*} ,  \cr
 \rho_{\lambda \lambda'}^{3} &=& 
 \frac{1}{N} \sum_{ \lambda_{\gamma}, \lambda_{i},\lambda_{f}}
\lambda_\gamma T_{\lambda_f , \lambda ; \lambda_i , \lambda_{\gamma}}
 T_{\lambda_f , \lambda' ; \lambda_i , \lambda_{\gamma}}^{*} ,    
\end{eqnarray}
where $\lambda_\gamma$, $\lambda_i$, and $\lambda_f$ represent the
helicities for the photon and the initial and final nucleons,
respectively, whereas $\lambda$ and $\lambda'$ denote those for the
$\phi$ meson. The normalization factor $N$ is defined as 
\begin{equation}
  \label{eq:app2}
  N = \sum|T_{\lambda_f,\lambda;\lambda_i, \lambda_{\gamma}}|^2.
\end{equation}

\end{document}